\PassOptionsToPackage{draft}{hypref}
\PassOptionsToPackage{breaklinks=true}{hypref}
\documentclass[a4paper,fleqn,usenatbib]{mnras}
\PassOptionsToPackage{draft}{hyperref}
\usepackage{graphicx}
\usepackage{amsmath}

\usepackage{longtable}

\usepackage{supertabular}

\makeatletter
\@fleqnfalse
\@mathmargin\@centering
\makeatother

\newcommand{\ProFound}{{\scshape ProFound}}
\newcommand{\DeepScan}{{\scshape DeepScan}}
\newcommand{\GALFIT}{{\scshape GALFIT}}

\newcommand{\Imfit}{{\scshape Imfit}}

\newcommand\uaermin{23}
\newcommand\uaermax{28}
\newcommand\rermin{0.3}
\newcommand\rermax{9.5}
\newcommand\nsources{175}  %With ignored targets removed
\newcommand\nudg{12}
\newcommand\priorrms{0.8}

\newcommand\nudgzerogc{0}
\newcommand\nallzerogc{12}
\newcommand\nallzerogcsig{106}

\newcommand\nudgrich{None}
\newcommand\nudgrichsig{5}
\newcommand\nlsbrichsig{13}

\newcommand\sigmaPF{2}
\newcommand\skymeshPF{100}
\newcommand\thresholdPF{1.03}
\newcommand\tolerancePF{1}
\newcommand\skycutPF{1}

\newcommand\Nsynthpersubframe{25000}
\newcommand\mminsynth{19}
\newcommand\mmaxsynth{26}

\newcommand\NspecGC{992}

\newcommand\ugmin{0.37}
\newcommand\ugmax{5.07}
\newcommand\grmax{1.23}
\newcommand\grmin{-0.18}
\newcommand\gimin{0.32}
\newcommand\gimax{2.00}

\newcommand\mugiblue{0.51}
\newcommand\mugired{1.12}
\newcommand\siggiblue{0.38}
\newcommand\siggired{0.40}
\newcommand\wgiblue{0.67}
\newcommand\wgired{0.33}
\newcommand\mugi{0.71}
\newcommand\siggi{0.48}

\author[D. J. Prole et al.]{D. J. Prole,$^{1,2}$\thanks{ProleD@cardiff.ac.uk}
									   M. Hilker,$^{1}$
									   R. F. J. van der Burg,$^{1}$
									   M. Cantiello,$^{3}$ 
									   A. Venhola,$^{4}$ \and
									   E. Iodice,$^{5}$ 
									   G. van de Ven,$^{1}$								   
									   C. Wittmann,$^{6}$ 
									   R. F. Peletier,$^{7}$ 
									   S. Mieske,$^{8}$  
									   M. Capaccioli,$^{5}$ \and
									   N. R. Napolitano,$^{5}$
									   M. Paolillo,$^{9}$
									   M. Spavone,$^{5}$
									   E. Valentijn$^{7}$
								   	   \\
								   	   $^{1}$European Southern Observatory, Karl-Schwarzschild-Str. 2, 85748 Garching bei M\"unchen, Germany \\
									  $^{2}$School of Physics and Astronomy, Cardiff University, The Parade, Cardiff, CF243AA, UK \\
									  $^{3}$INAF–Osservatorio Astronomico d'Abruzzo, Via M. Maggini snc, 64100, Teramo, Italy\\
									  $^{4}$Division of Astronomy, Department of Physics, University of Oulu, Oulu, Finland\\
						              $^{5}$INAF-Astronomical Observatory of Capodimonte, via Moiariello 16, Naples, I-80131, Italy\\
									  $^{6}$Astronomisches Rechen-Institut, Zentrum f\"ur Astronomie der Universit\"at Heidelberg, M\"onchhofstra\ss e 12-14, 69120 Heidelberg, Germany \\
									  $^{7}$Kapteyn Astronomical Institute, University of Groningen, PO Box 72, 9700 AV Groningen, The Netherlands\\
									  $^{8}$European Southern Observatory, Alonso de Cordova 3107, Vitacura, Chile\\
									  $^{9}$University of Naples “Federico II”, C.U. Monte Sant’Angelo, Via Cinthia, 80126, Naples, Italy
								  	  }

\title[Halo mass estimates of LSB galaxies]{Halo mass estimates from the Globular Cluster populations of \nsources\ Low Surface Brightness Galaxies in the Fornax Cluster}

\date{Accepted XXX. Received YYY; in original form ZZZ}

\pubyear{2018}

\begin{document}
	\label{firstpage}
	\pagerange{\pageref{firstpage}--\pageref{lastpage}}
	\maketitle

\begin{abstract}
	
The halo masses $M_{halo}$ of low surface brightness (LSB) galaxies are critical measurements for understanding their formation processes. One promising method to estimate a galaxy's $M_{halo}$ is to exploit the empirical scaling relation between $M_{halo}$ and the number of associated globular clusters ($N_{\mathrm{GC}}$). We use a Bayesian mixture model approach to measure $N_{\mathrm{GC}}$ for \nsources\ LSB ($\uaermin\leq\left\langle \mu_{e,r} \right\rangle [\mathrm{mag\ arcsec}^{-2}]\leq\uaermax$) galaxies  in the Fornax cluster using the Fornax Deep Survey (FDS) data; this is the largest sample of low mass galaxies so-far analysed for this kind of study. The proximity of the Fornax cluster  means that we can measure galaxies with much smaller physical sizes ($\rermin\leq r_{e,r}\ [\mathrm{kpc}]\leq\rermax$) compared to previous studies of the GC systems of LSB galaxies, probing stellar masses down to $M_{*}\sim10^{5}\mathrm{M_{\odot}}$. The sample also includes \nudg\ ultra-diffuse galaxies (UDGs), with projected $r$-band half-light radii greater than 1.5 kpc. Our results are consistent with an extrapolation of the $M_{*}-M_{halo}$ relation predicted from abundance matching. In particular, our UDG measurements are consistent with dwarf sized halos, having typical masses between $10^{10}$ and $10^{11}\mathrm{M_{\odot}}$. Overall, our UDG sample is statistically indistinguishable from smaller LSB galaxies in the same magnitude range. We do not find any candidates likely to be as rich as some of those found in the Coma cluster. We suggest that environment might play a role in producing GC-rich LSB galaxies.

\end{abstract}


\section{Introduction}

Low surface brightness (LSB) galaxies are among the most common in the Universe, yet observational challenges \citep{Disney1976} mean that they are also among the most mysterious. The existence of large LSB galaxies is a well known phenomenon. They were first detected several decades ago \citep[e.g.][]{Sandage1984, Bothun1987, Impey1988} but have received renewed interest in more recent years. However, their intrinsic properties and formation histories are still not fully understood and it is not clear whether they represent a distinct population from smaller LSB dwarf galaxies, which can form naturally in high spin halos expected from hierarchical galaxy formation models \citep[e.g.][]{Dalcanton1995, Jimenez1998} and from harassment of normal dwarf galaxies \citep{Moore1998, Mastropietro2005}. 

\indent An outstanding question is whether there is truly anything unique about the way in which large LSB galaxies form in comparison to their smaller counter-parts, given that they seem to share a continuous distribution of observable properties \citep{Conselice2003, Wittmann2017, Conselice2018}. Since \cite{VanDokkum2015} detected a surprisingly high abundance of large LSB galaxies (that they termed ultra-diffuse galaxies or UDGs) in the Coma Cluster, numerous theories have been proposed to explain their origins. Initially,  \cite{VanDokkum2015} suggested they could reside in massive halos, similar in total mass to the Milky Way with a truncated star formation history. This is referred to as the ``failed L$\star$'' scenario, and is supported observationally by their unusually large sizes (optical effective radii $\geq1.5$ kpc) together with their LSB ($\mu_{0}^{g}\geq\sim23$), red colours and abundance in dense environments.

\indent There are several possible mechanisms to explain the existence of UDGs other than the failed L$\star$  scenario. \cite{Yozin2015} have shown that ram-pressure stripping resulting from an early in-fall to cluster environments is sufficient to reproduce several properties of the Coma UDGs. Other authors have shown that different environmental effects like tidal heating may be enough to explain their formation \citep{Collins2013, Carleton2018}. While these models may seem to indicate that UDGs are phenomena associated preferentially with dense environments \cite[supported observationally by][]{VanderBurg2017}, it is also thought that a field population should exist \citep{McGaugh1996, DiCintio2017}, plausibly arising from the high angular momentum tail of the dwarf galaxy population \citep{Amorisco2016} or from secular evolution processes such as supernovae feedback. Of course, there could be multiple formation scenarios for UDGs that combine both secular and environmentally-driven processes \citep{Jiang2018}. 

\indent The halo mass is a key parameter in distinguishing between formation models of UDGs. Typically, current models favour dwarf-sized halos with truncated star formation histories \citep[e.g.][]{Rong2017, Amorisco2016}, making them similar to normal LSB galaxies but larger. UDGs are abundant in high density environments such as in the centres of clusters \citep[e.g.][]{Mihos2015, Koda2015, Venhola2017} where they require a relatively high dark matter fraction in order to survive. However, it is not clear whether UDGs can form with lower mass-to-light ratios (M/L) in less dense environments such as the field \citep{VanDokkum2018, Trujillo2018}.

\indent There have been several attempts to constrain the halo masses of UDGs with a variety of measurement techniques used, mainly focussing on UDGs in groups and clusters. Metrics include weak lensing \citep{Sifon2018}, prevalence of tidal features as a function of cluster radius \citep{Mowla2017}, comparisons of their spatial distribution with that of dwarf and massive galaxies \citep{vanderBurg2016, Roman2017}, richness of their globular cluster systems \citep{Beasley2016b, Amorisco2018, VanDokkum2017, Lim2018} as well as direct measurements of the velocity dispersions of stellar populations \citep{VanDokkum2016} and globular cluster systems \citep{Beasley2016a, Toloba2018}.

\indent Globular clusters offer an interesting insight into the formation mechanisms of LSB galaxies. They are thought to form mainly in the early epochs of star formation within massive, dense giant molecular clouds that are able to survive feedback processes that might otherwise shut off star formation in their host galaxy  \citep{Hudson2014, Harris2017}. The halo mass of galaxies has been shown to correlate well with both the number of associated GCs ($N_{\mathrm{GC}}$) and the total mass of their GC systems \citep[$M_{\mathrm{GC}}$; e.g.][]{Spitler2009, Harris2013, Harris2017}, which means measurements of either $N_{\mathrm{GC}}$ or $M_{\mathrm{GC}}$ can be used to constrain $M_{halo}$. However, \cite{Forbes2018} show that the traditional relation between $N_{\mathrm{GC}}$ and $M_{halo}$ may lose accuracy in the low $M_{halo}$ regime, perhaps because lower mass galaxies tend to have lower mass GCs without a common mean GC mass. Additionally, it has been shown that there is a correlation between the GC half-count radius and $M_{halo}$ \citep{Forbes2017,  Hudson2018}.

\indent The majority of studies of the GC populations of UDGs have up until now focussed on the Coma galaxy cluster, the most massive \cite[$M_{\mathrm{tot}}\sim6\times10^{14}\mathrm{M_{\odot}}$,][]{Hughes1998} galaxy cluster within 100 Mpc. In this paper we analyse exclusively galaxies in the core of the Fornax cluster. In comparison to Coma, it is around five times closer \citep[$d\sim$20 Mpc,][]{Blakeslee2009} but less massive \citep[$M_{\mathrm{tot}}\sim7\times10^{13}M_{\odot}$,][]{Drinkwater2001}. Using the empirical relation of \cite{VanderBurg2017}, there are approximately 10 times less UDGs expected in Fornax than in Coma, many of which have been catalogued already \citep{Munoz2015, Venhola2017}.

\indent While overall we have a relatively small sample of UDGs, an advantage of working with the Fornax cluster is that cluster members have much larger projected sizes compared to the background galaxy population, so we can analyse the population of smaller LSB galaxies at the same time as the UDG population without contamination from interlopers. Indeed, much of the new literature surrounding LSB galaxies focusses on UDGs and this may be in-part due to the relative ease of distinguishing larger galaxies from background objects in group or cluster environments. A second advantage of Fornax over Coma is that GCs are brighter in apparent magnitude by $\sim$3.5 mag due to their relative proximity, meaning that we can probe further into the GC luminosity function.

\indent We note that the relatively large number of galaxies we analyse in this study is important for at least partially overcoming systematic uncertainties involved in measuring halo masses with low numbers of tracers as made clear by \cite{Laporte2018} and the possible stochastic nature of the $M_{*}-M_{halo}$ relation at low mass \citep{Brook14, Errani2018}.

\indent In this work we provide constraints on the halo masses for a selection of LSB galaxies first identified by \cite{Venhola2017} using the optical Fornax Deep Survey \citep[FDS,][]{Iodice2016}. The structure of the paper is as follows: We describe the data in $\S$\ref{section:data}. In $\S$\ref{section:method}  we describe the method to detect globular cluster candidates (GCCs) and infer the total number of GCs associated with our target galaxies. We provide our results in $\S$\ref{section:results}, where we estimate the halo masses from the inferences on $N_{\mathrm{GC}}$ and $M_{\mathrm{GC}}$ using the empirical scaling relations of \cite{Harris2017}. We discuss our results and provide conclusive remarks in $\S$\ref{section:conclusion}. We use the AB magnitude system throughout the paper, and adopt a distance of 20Mpc to the Fornax cluster.

 %\indent In this work we provide constraints on the halo masses for a selection of low surface brightness galaxies (LSBGs) first identified by \cite{Venhola2017} using the optical Fornax Deep Survey \citep[FDS,][]{Iodice2016}. In brief, this is accomplished by 1) subtracting models of the sources from the data using \Imfit\  \citep{Erwin2015}; 2) Creating a catalogue of GC candidates using the {\tt ProFound}\footnote{https://github.com/asgr/ProFound} package \citep{Robotham2018} for photometry; 3) Applying a colour selection based on a set of confirmed Fornax GCs in a similar fashion to \cite{DAbrusco2016}; 4) Statistically inferring $N_{\mathrm{GC}}$ using a Bayesian mixture model approach similar to that of \cite{Amorisco2018}; 5) Correcting for the detection efficiency based on synthetic source injections and literature measurements of the GC luminosity function; 6) Inferring the halo masses of the galaxies using the empirical relation of \cite{Harris2017}.
 
 %%%%%%%%%%%%%%%%%%%%%%%%%%%%%%%%%%%%%%%%%%%%%%%%%%%%%%%%%%%%%%%%%% %%%%%%%%%%%%%%%%%%%%%%%%%%%%%%%%%%%%%%%%%%%%%%%%%%%%%%%%%%%%%%%%%% 

\section{Data}
\label{section:data}

We use the four central 1$\times$1 degree$^{2}$ frames of the FDS (FDS IDs 10, 11, 12 \& 16), i.e. the same region used by \cite{Venhola2017} in their by-eye classification of low surface brightness sources in the Fornax galaxy cluster. These data were obtained using the OmegaCAM \citep{Kuijken2012} instrument on the 2.6m ESO VLT Survey Telescope \citep[VST,][]{Capaccioli2012} in the $u'$, $g'$, $r'$ \& $i'$ bands. We note that Fornax GCs are unresolved in our data such that we consider them as point sources throughout the paper.

\indent We specifically used the VSTtube-reduced FDS data \citep{Grado2012, Capaccioli2015}, which is optimised for point-source photometry but is not as deep as the data used by \cite{Venhola2017}, which is reduced using a combination of the OmegaCAM pipeline and AstroWISE \citep{McFarland2011}, but with a slightly wider PSF than the VST-tube reduction because images with poor seeing were included in the stacks. We also performed additional photometric corrections to bring our photometry into the AB magnitude system as described in appendix \ref{section:photcal}.

%%%%%%%%%%%%%%%%%%%%%%%%%%%%%%%%%%%%%%%%%%%%%%%%%%%%%%%%%%%%%%%%%% 
%%%%%%%%%%%%%%%%%%%%%%%%%%%%%%%%%%%%%%%%%%%%%%%%%%%%%%%%%%%%%%%%%% 

\section{Methodology}
\label{section:method}

In this work we target the GC populations of galaxies identified by-eye in the \cite[][hereafter V17]{Venhola2017} catalogue. We split the sample into two groups: Low surface brightness galaxies (LSBGs), defined as those with $r$-band effective radii $r_{e,r}<1.5$ kpc and UDGs, defined as those with $r_{e,r}\geq 1.5$ kpc. The sources are defined as LSB because they were measured to have central surface brightness $\mu_{0}^{r}\geq23$ by V17. We omit two UDGs (FDS11\_LSB1 and FDS11\_LSB17) from the sample because they are in significantly crowded locations and measuring their properties accurately would require a more sophisticated analysis. 

\begin{table}
	\begin{tabular}{| c | c | }
		\hline 
		Parameter & Constraint \\
		\hline 
		{\tt mag} ($g$) & 14 to 19 [mag]\\
		{\tt axrat} & $\geq0.95$ \\
		{\tt Nobject} & 0 \\
		{\tt Nmask} & 0 \\
		\hline 
	\end{tabular}
	\caption{\ProFound\ measurement constraints for point source selection in the PSF modelling. {\tt axrat} is the axis-ratio. {\tt Nobject} is the number of pixels belonging to the segment that are touching another source. {\tt Nmask} is the number of pixels belonging to the segment that are touching a masked region.  See the \ProFound\ documentation for more details of these parameters. Further criteria are discussed in the text.}
	\label{table:psf}
\end{table}

\indent Before running our detection algorithm, we subtract model galaxy profiles in each band using \Imfit\  \citep[][see Appendix \ref{section:imfit}]{Erwin2015}. We were unable to get a stable \Imfit\ model for three sources (FDS11\_LSB16, FDS12\_42, FDS12\_47) because they were too faint, so we adopt the measurements of V17 (made from deeper stacks) for these sources and rely on a separate background subtraction procedure to remove the galaxy light (see $\S$\ref{section:colourmeas}). We select only galaxies with measured $r$-band effective radii greater than 3$\arcsec$ ($\sim$ 0.3kpc at Fornax distance) so that we target cluster members with confidence \citep{Sabatini2003, Davies2016}.

\indent We used the  \ProFound\footnote{https://github.com/asgr/ProFound} package \citep{Robotham2018} for the source detection and photometry, with the following settings: {\tt skymesh=\skymeshPF} pixels ($\sim$20\arcsec), {\tt sigma=\sigmaPF} pixels, {\tt threshold=\thresholdPF}, {\tt tolerance=\tolerancePF}, {\tt skycut=\skycutPF}. All other settings were defaults. Our detection was performed exclusively on the $g$-band (the deepest) so that we could easily measure and account for our detection efficiency without considering the colours of individual sources. We note that we split the four FDS frames into $9\times9$ subframes to ease the memory requirements for \ProFound.

%%%%%%%%%%%%%%%%%%%%%%%%%%%%%%%%%%%%%%%%%%%%%%%%%%%%%%%%%%%%%%%%%% 

\subsection{PSF models}
\label{section:psf}

\indent We obtained point spread function (PSF) models for each band and subframe using our \ProFound\ measurements as follows. Bright, unsaturated point sources were selected  in the \ProFound\ {\tt mag} - {\tt R50} (approx. half-light radius) plane, using the selection criteria listed in table \ref{table:psf}. Additionally we sigma-clipped the measurements in {\tt R50} (approximately flat over the magnitude range for point sources) and offset the relation by 4$\sigma$ with respect to the median to measure an upper-limit on {\tt R50} for the selection. 

\begin{figure}
	\includegraphics[width=\linewidth]{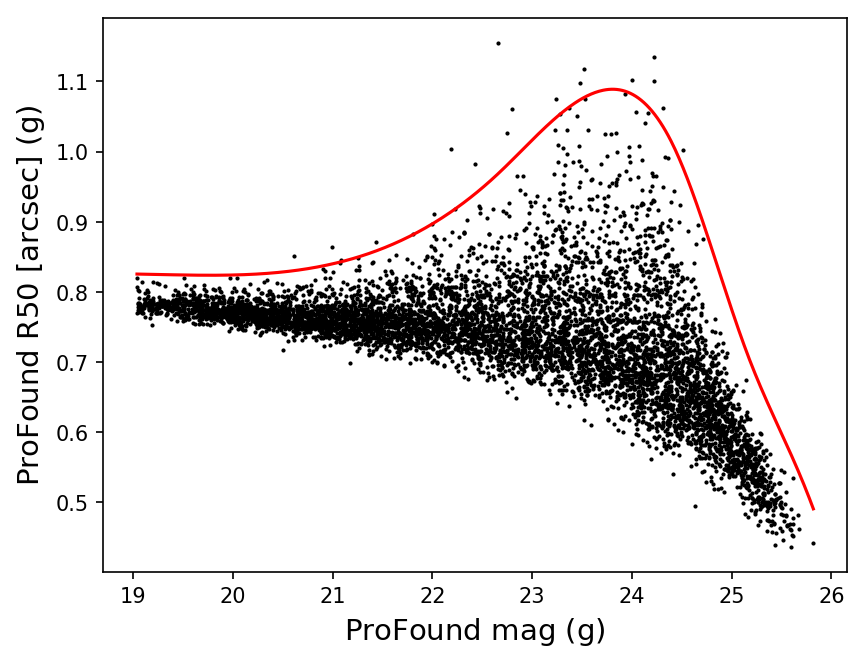}
	\centering
	\caption{Point source selection function for a subframe (red line) obtained by fitting a cubic spline to measurements from the synthetic source injections (black points) after offsetting {\tt R50} by $4\sigma$ in bins of magnitude with respect to the median. All points lower than the red line are selected. The increase in scatter towards the faint end means that a simple cut in {\tt R50} would either result in a low recovery efficiency or high level of contamination from our point source selection.}
	\label{figure:point}
\end{figure}

\indent We used \Imfit\  to fit a model Moffat profile (keeping the axis-ratio as a free parameter) to each point source following a local sky subtraction. We did not stack individual point source cut-outs to avoid artificial widening of the PSF caused by misalignment of the images. The resulting distribution of model Moffat fits was then sigma-clipped at 3$\sigma$ in the FWHM-concentration index plane to remove outliers caused by bad fits. We finally selected a fiducial model PSF for each band and subframe by adopting the fit with the average FWHM. 

\indent Of primary importance for our analysis are the $g$-band PSF models. While for a specific FDS frame we found little variation of the PSF over its subframes, on a frame-by-frame basis the \Imfit\ FWHM ranges between approximately 0.7 and 1.2$\arcsec$. 

%%%%%%%%%%%%%%%%%%%%%%%%%%%%%%%%%%%%%%%%%%%%%%%%%%%%%%%%%%%%%%%%%%

\subsection{Point source selection}
\label{section:pointsel}

\indent We used synthetic source injections based on our Moffat PSF models from $\S$\ref{section:psf} to produce our point source selection function and quantify our recovery efficiency (RE). We injected $\sim$\Nsynthpersubframe\ synthetic profiles per subframe into the real data at random locations in the vicinities ($420\times420\arcsec$ cut-outs) of our target galaxies after subtracting the galaxy models from the data. This was done in the $g$-band, with apparent magnitudes ranging between \mminsynth\ and \mmaxsynth. Our matching criteria for the synthetic sources was that the central coordinate of the injected source had to lie on top of a segment in the  \ProFound\ segmentation map. Additionally, we only considered sources that did not match with segments from the result of running  \ProFound\ over the original frames (i.e. without the injected sources). We note that the measurements of the synthetic point sources are in good agreement with measurements of real sources when plotted on the {\tt mag-R50} plane. 

\indent Once we had acquired the \ProFound\ measurements of the synthetic sources, we fitted a smooth cubic spline to the data in the \ProFound\ {\tt mag} and {\tt R50} plane. Specifically, the spline was fit to the data binned in {\tt mag}, and positively offset by 4$\sigma$ of the {\tt R50} values within the bin. See figure \ref{figure:point}. The rationale behind this was that as sources become fainter, the scatter in {\tt R50} increases such that a simple cut at a specific value would either be too high for bright objects or conversely too low for some of the fainter objects with large values of {\tt R50}. We obtained a different point source selection function for each subframe. 

%%%%%%%%%%%%%%%%%%%%%%%%%%%%%%%%%%%%%%%%%%%%%%%%%%%%%%%%%%%%%%%%%% 

\subsection{Colour measurement}
\label{section:colourmeas}

\begin{figure}
	\includegraphics[width=\linewidth]{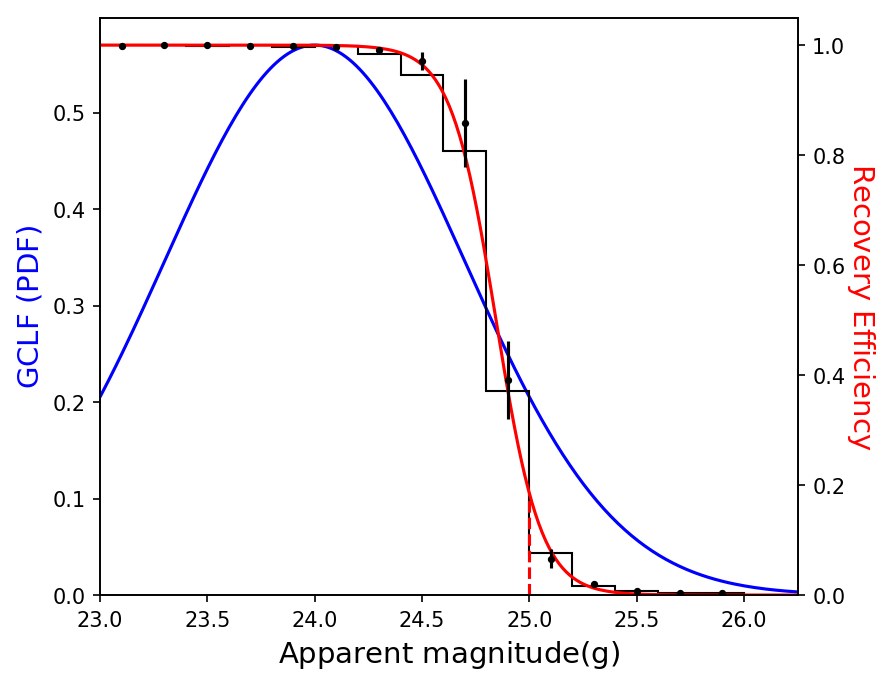}
	\centering
	\caption{$g$-band recovery efficiency for a subframe of injected point sources (black histogram) after applying our selection criteria, along with the RE (red line) and a fiducial GCLF (blue line). The dashed red line shows our additional cut at $m_{g}$=25. The GCLF turnover magnitude is clearly reached by our detection pipeline. Integrating the recovery efficiency over the GCLF (including the magnitude cuts) yields an overall recovery efficiency of $\sim80\%$. Also shown is the mean and standard deviation of the RE across all the subframes for FDS frame 10 (black errorbars).}
	\label{figure:RE}
\end{figure}

\indent We obtained aperture magnitudes of the \ProFound\ sources in fixed apertures of diameter 5 pixels in all the bands. The sky level and its uncertainty were calculated for each detected source by placing many identical apertures in 51$\times$51 pixel cut-outs (the fiducial FWHM of 1$\arcsec$ is $\sim5$ pixels) and recording the median and standard-deviation of the contained flux values after sigma clipping these at 2$\sigma$ to remove contamination from other sources. Additionally the sky apertures were placed at radii greater than 20 pixels from the centre of the source.

\indent These magnitudes were then corrected for the PSF size in each band through calibration against an existing catalogue of PSF-corrected point sources made using the same data (Cantiello et al.; in prep.). See appendix \ref{section:photcal} for a discussion of the photometric calibration procedure. We note that we have not used this catalogue for this work because of the need to subtract galaxy profiles from the data and the need to quantify the RE.

%%%%%%%%%%%%%%%%%%%%%%%%%%%%%%%%%%%%%%%%%%%%%%%%%%%%%%%%%%%%%%%%%% 

\subsection{Recovery Efficiency}
\label{section:RE}

\indent We quantified the RE separately for each subframe using the point source selection functions with the synthetic source measurements. We imposed a faint-end limit on the corrected $g$-band aperture magnitude of 25 mag because measuring accurate colours at fainter magnitudes is more difficult and because the degeneracy between point sources and other faint sources in the {\tt mag-R50} plane is exacerbated in this region.  Additionally, we apply a lower-bound cut in the corrected $g$-band aperture magnitude of 21 mag to reduce possible contamination from bright stars, ultra-compact dwarf galaxies (UCDs) and nuclear star clusters (NSCs).  

\indent The RE itself was measured by taking the ratio of detected and selected point source injections to the total number of injected sources in bins of intrinsic magnitude. A sigmoid function,

\begin{equation}
\epsilon(m) = [1+\exp{(-k_{1}(k_{2}-m)})]^{-1}
\label{equation:sigmoid}
\end{equation}

\noindent was fit to the result (see figure \ref{figure:RE}). The recovery efficiency $\epsilon(m)$ is sufficient to reach the turnover magnitude of the $g$-band GC luminosity function (GCLF), which is approximated by a Gaussian function centred at 24$m_{g}$ at the distance of Fornax \citep{Villegas2010}. We adopt a value of 0.7 for the GCLF standard deviation, which is a reasonable estimate for low surface brightness galaxies \citep{Trujillo2018}. Under these assumptions, our estimated GC completeness ranges between $\sim$60\% and 90\% depending on the subframe. The mean completeness is estimated to be 82\% across all the subframes. Of course, this number depends on the exact form of the adopted GCLF. While its peak at 24$m_{g}$ is fairly well known \citep[The peak of the GCLF can sometimes be used as a standard candle, see][]{Rejkuba2012}, a degree of uncertainty is attributed to its width. We discuss the effects of varying the GCLF on our results in $\S$\ref{section:GCLF}.

%%%%%%%%%%%%%%%%%%%%%%%%%%%%%%%%%%%%%%%%%%%%%%%%%%%%%%%%%%%%%%%%%% 

\subsection{Colour Selection}
\label{section:coloursel}

\indent We have applied a colour selection to our point sources to produce a catalogue of globular cluster candidates (GCCs) for each target galaxy, using as few assumptions about the underlying GC colours as possible. The full colour space in $u, g, r, i$ was used for the selection. This is important because of the need to remove interloping point sources from our final GCC catalogue, which include foreground stars and unresolved background galaxies. However, we point out that both interloping populations are partially degenerate in colour space with the actual GCs \citep[see also][]{Pota2018} and these sources must be accounted for using spatial information (see $\S$\ref{section:bayes}).

\indent This colour selection was accomplished by first cross-matching our point sources from the four FDS frames with a compilation of spectroscopically confirmed Fornax compact objects \citep{Schuberth2010, Wittmann2016, Pota2018}. This resulted in a catalogue of \NspecGC\ matching sources. We note that we partially account for bright UCDs with our bright-end magnitude cut-off. The external catalogue has a magnitude distribution that drops off quickly at magnitudes fainter than $\sim23 m_{g}$ and so is not complete for our purposes and this limited depth has to be accounted for.

\begin{figure}
	\includegraphics[width=\linewidth]{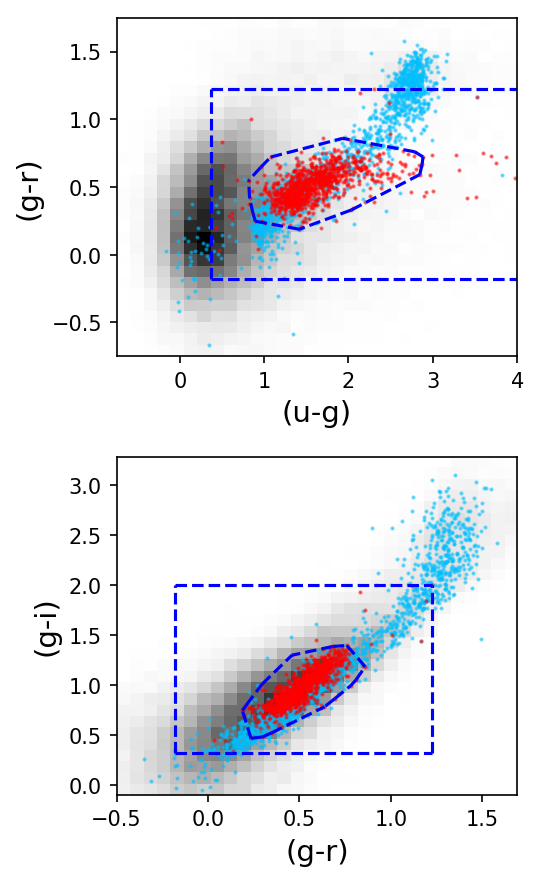}
	\centering
	\caption{Colour-colour measurements of spectroscopic Fornax compact objects (red points), galactic stars (blue points) against the empirical distribution of all detected point sources over a subframe (greyscale histogram). Also shown are two colour selection boxes (blue dashed lines). In each panel, the large box corresponds to the minimum-bounding rectangle of compact object measurements in this colour-colour plane, and the smaller box is produced using the {\tt DBSCAN} algorithm.}
	\label{figure:colsel}
\end{figure}

\indent We used the density-based clustering algorithm {\tt DBSCAN} \citep{Ester1996} to define regions in the $(u-g)-(g-r)$ and $(g-r)-(g-i)$ planes separately for our GCC selection. We used a clustering radius of 0.1 mag and required at least 5 spectroscopic GCs within this radius for clusters to form. After acquiring the DBSCAN clusters, we fitted a convex hull to all the clustered points and used this as the boundary of the selection box; the results of this are shown in figure \ref{figure:colsel}. Approximately 93\% of the spectroscopic GCs occupy the selection region and we correct for this factor in our later inferences on $N_{\mathrm{GC}}$.

\begin{figure}
	\includegraphics[width=\linewidth]{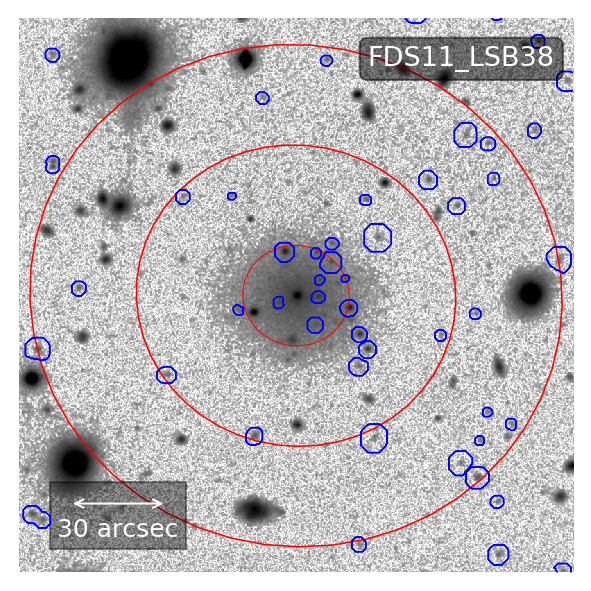}
	\centering
	\caption{Result of full GCC identification and selection for one of our target galaxies. The selected sources are shown using their \ProFound\ segments in blue. We also show the 1, 3 and 5 $r_{e,r}$ contours from the \Imfit\ modelling. A clear over-density of sources can be seen close to the centre of the galaxy.}
	\label{figure:demo}
\end{figure}

\indent The fraction of GCs that occupy the colour selection box decreases as a function of magnitude because of measurement error. Thus, we have used a probabilistic approach to identify all sources that \textit{could} occupy the box, given their uncertainty. Specifically, we selected all sources that were consistent within $2\sigma$ of their measurement uncertainty of the box, separately in each colour-colour plane. While the colour selection box was measured in magnitude units, we actually converted it into linear flux-ratio units (accounting for the photometric calibration) to select GCCs. This was done primarily to overcome the effects of the shallow $u$-band which would otherwise impact our estimate of the RE. A visual example of our combined point source selection criteria with colour selection for one of our target galaxies is shown in figure \ref{figure:demo}.

\indent We also performed a separate analysis using a much wider colour-selection box, also shown in figure \ref{figure:colsel}. We measured the minimum bounding rectangles in each colour-colour plane from the matching sources, forming a 3D colour selection box. The box is bounded by \grmin$<(g-r)<$\grmax, \gimin$<(g-i)<$\gimax, \ugmin$<(u-g)<$\ugmax; the high upper-limit on $(u-g)$ is likely due to scatter caused by the shallow $u$-band. While conservative in nature, the box is sufficient to contain all the matching spectroscopic GCs down to $m_{g}\sim23$. We note here that our overall results are not significantly impacted by this change. We refer to the results obtained using the {\tt DBSCAN} colour box for the remainder of the paper.

%\indent While this selection contains all the matching spectroscopic GCs, it is not guaranteed that all the GCs in our sample occupy it because of measurement uncertainty (particularly for faint sources). Thus, we have used a probabilistic approach to identify all sources that \textit{could} occupy the box, given their uncertainty. Specifically, we selected all sources that we cannot exclude from the box at a confidence level of greater than 99\%, assuming Gaussian errors in colour space. The result of this selection is that essentially all the GCs should be identified in colour space; we therefore assume a negligible effect on the RE due to the colour selection. A visual example of our combined point source selection criteria with colour selection for one of our target galaxies is shown in figure \ref{figure:demo}.

\indent We note that we do not fit for the intrinsic colour distribution of GCs and interlopers as was done in \cite{Amorisco2018}. The reason for this is that simple statistical representations (e.g. Gaussian) are inappropriate to describe our data in the multi-dimensional colour space. This can be gathered from the appearance of figure \ref{figure:colsel}. It may be possible to include extra colour-terms in the mixture models described in $\S\ref{section:bayes}$, but we leave this for future work.

%%%%%%%%%%%%%%%%%%%%%%%%%%%%%%%%%%%%%%%%%%%%%%%%%%%%%%%%%%%%%%%%%% 

\subsection{Bayesian Mixture Models}
\label{section:bayes}

We adopt a simplified version of the Bayesian mixture modelling of \cite{Amorisco2018} to measure the properties of the GC systems of our target galaxies. We are similarly motivated to rescale the spatial coordinates of the GCCs into units of the 1$r_{e,r}$ (half-light radius of the galaxy) ellipse. Our model consists of two surface density components: A central Plummer profile to represent the GCs associated with the target galaxy,

\begin{equation}
\mathrm{\Sigma}(r, r_{h}) = \frac{1}{\pi}\frac{1}{r_{h}^{2}(1+r^{2}/r^{2}_{h})^{2}}
\label{equation:plummer}
\end{equation}

\noindent where $r_{h}$ is the half number radius, in units of $r_{e,r}$, and a uniform distribution to represent the background, which mainly consists of stars, background galaxies and intra-cluster GCs. The presence of NGC1399 in the centre of Fornax means that its GC system may contribute to a non-uniform background in its vicinity. However, it can be shown that for our galaxies the gradient in the surface density of GCs belonging to NGC1399 is negligible, with a maximal gradient value of $\sim10^{-4}$ objects arcmin$^{-3}$ in the vicinity of our sources. For this calculation, we have used the de Vaucouleurs' fit to the GC system of NGC1399 from \cite{Bassino2006}. The total model likelihood takes the form:

\begin{equation}
\mathcal{L}=\prod_{i=1}^{N_{\mathrm{GCC}}}\left[\frac{f\Sigma(r_{i},r_{h})}{\int_{0}^{r_{\mathrm{max}}}S(r')\Sigma(r',r_{h})r'dr'}+\frac{1-f}{\int_{0}^{r_{\mathrm{max}}}S(r')r'dr'}\right]
\end{equation}

\noindent where $i$ runs over all detected (and not masked) GCCs within the transformed radius $r_{\mathrm{max}}$ of the galaxy, which we fix as 15$r_{e,r}$; large enough to include all the galaxies' GCs and a large number of background GCCs. We do not consider larger regions because of the increased potential of contamination from steep GCC gradients in the Fornax core. The spatial completeness function $S(r)$ encodes the fractional unmasked area as a function of radius. There are two free parameters: $f$, the mixing fraction (i.e. the fraction of all sources that are GCs belonging to the target galaxy) and the ratio $r_{h}$/$r_{e}$. We do not explicitly include morphological or colour terms in the model likelihood, but account for this in the GCC selection described in sections \ref{section:pointsel} and \ref{section:coloursel}.

\indent We impose a Gaussian prior on the ratio $r_{h}$/$r_{e}$ based on the results of \cite{Amorisco2018}. The prior is centred at $\sim1.5r_{e}$ with a standard deviation of \priorrms\ and truncated at zero. The choice of prior is very influential, particularly in the low $f$ regime in which most of our sources are anticipated to lie. However, since \cite{Amorisco2018} probe a similar sample of sources in a similar environment (the Coma cluster) to ours and that the  $r_{h}$/$r_{e}\simeq1.5r_{e}$ relationship appears elsewhere in the literature \citep{VanDokkum2017, Lim2018} it is a reasonable estimate. We probe the effects of modifying the prior on $r_{h}$ in appendix \ref{section:prior}. The prior width is much greater than the RMS of the median values quoted by \cite{Amorisco2018}, so that if there is any significant deviation it should be recognised in our analysis.

\begin{figure*}
	\includegraphics[width=\linewidth]{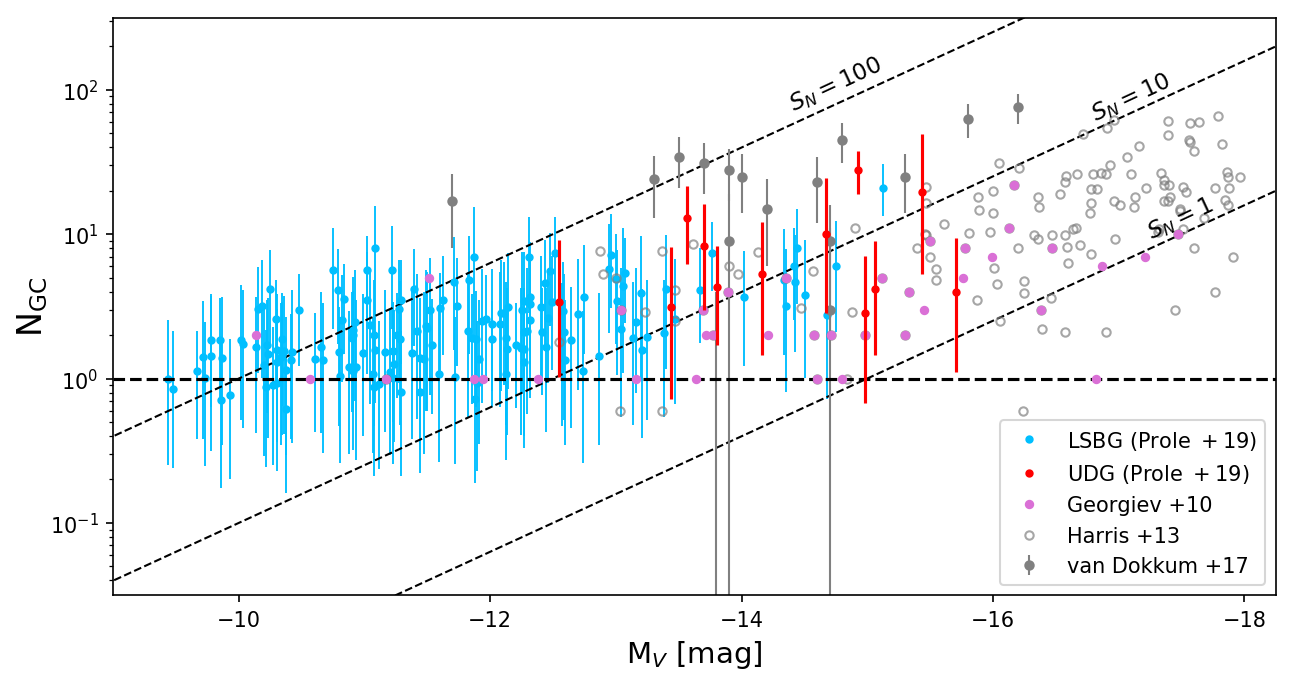}
	\centering
	\caption{Corrected number counts of globular clusters as a function of absolute $V$-band magnitude. On the figure we show our new measurements in blue (LSBG sample) and red (UDG sample). The points represent median values from our MCMC posterior, and the errorbars span the range of the 1$\sigma$ uncertainties. The open grey circles belong to the \protect\cite{Harris2013} catalogue. The grey points with error bars are the sample of selected Coma UDGs from \protect\cite{VanDokkum2017}. We also plot the sample of dwarf galaxies measured by \protect\cite{Georgiev2010} in pink. The diagonal dashed black lines represent $S_{N}$ contours, whereas the horizontal one represents 1 GC.}
	\label{figure:Ngc}
\end{figure*}

%%%%%%%%%%%%%%%%%%%%%%%%%%%%%%%%%%%%%%%%%%%%%%%%%%%%%%%%%%%%%%%%%% %%%%%%%%%%%%%%%%%%%%%%%%%%%%%%%%%%%%%%%%%%%%%%%%%%%%%%%%%%%%%%%%%% 

\section{Results}
\label{section:results}

\subsection{Inference on Globular Cluster numbers}
\label{section:ngc}

\indent We made data cut-outs in each band for each source that were 15$\times$15 $r_{e,r}$ in size. We chose this size because tests with mock datasets (with realistic numbers of interlopers derived from the data) revealed that the measured number of GCs was negatively biased for much smaller values, and this particularly affected systems with less than 10 intrinsic GCs. At 15$\times$15 $r_{e,r}$, we were able to recover unbiased measurements of $N_{\mathrm{GC}}$ even for systems with no GCs. 

\indent The GCCs were selected according to the criteria described in $\S$\ref{section:pointsel} and by their colour, described in $\S$\ref{section:coloursel}. All non-selected sources were masked using their \ProFound\ segments. We additionally automatically masked the areas around \ProFound\ sources with $g$-band magnitudes brighter than 19 in an effort to remove interloping GCs belonging to other systems. This was accomplished by placing elliptical masks scaled to 2 times the \ProFound\ {\tt R100} radius.

\indent All sources in the GCC catalogues that had central coordinates overlapping with the masks were removed. The spatial completeness function could then be measured by measuring the masked fraction in concentric annuli centred on the galaxy, spaced by 0.01$r_{e,r}$ and linearly interpolating the result. We note that two sources\footnote{FDS10\_LSB33, FDS11\_LSB32} were omitted from the analysis because they were almost completely masked.

\indent We then ran the Monte-Carlo Markov chain (MCMC) code {\scshape emcee}\footnote{http://dfm.io/emcee/current/} to obtain the posterior distributions of $f$ and $r_{h}$ for each individual target galaxy. The final inference on the number of GCs associated with each galaxy, $N_{\mathrm{GC}}$, was calculated as

\begin{equation}
N_{\mathrm{GC}}^{j} = f^{j}\frac{\int_{0}^{r_{\mathrm{max}}}\Sigma(r,r_{h}^{j})rdr}{\int_{0}^{r_{\mathrm{max}}}S(r)\Sigma(r,r_{h}^{j})rdr}\frac{\int_{m_{\mathrm{1}}}^{m_{\mathrm{2}}}g(m)dm}{\int_{m_{\mathrm{1}}}^{m_{\mathrm{2}}}\epsilon(m)g(m)dm}N_{\mathrm{GCC}}
\end{equation}

\noindent taking into account the masked fraction and magnitude incompleteness. Here, $j$ indicates the posterior index and $g(m)$ is the Gaussian $g$-band GCLF. The results of this are shown in figure \ref{figure:Ngc}, where we convert our galaxy photometry to $V$-band magnitudes using the prescriptions of \cite{Jester2005}. As a means of comparison, we show in Appendix \ref{section:ML} that our inferences on $N_{\mathrm{GC}}$ are consistent with the measurements of \cite{Miller2007} for a small sample of overlapping galaxies using a chi-squared test. 

\iffalse%%%%%%%%%%%%%%%%%%%%%%%
\begin{figure*}
	\includegraphics[width=\linewidth]{mv_ngc_mcmc2.png}
	\centering
	\caption{Corrected number counts of globular clusters as a function of absolute $V$-band magnitude. On the figure we show our new measurements in blue (LSBG sample) and red (UDG sample). The points represent median values from our MCMC posterior, and the errorbars span the range of the 1$\sigma$ uncertainties. The open grey circles belong to the \protect\cite{Harris2013} catalogue. The grey points with error bars are the sample of selected Coma UDGs from \protect\cite{VanDokkum2017}. We also plot the sample of dwarf galaxies measured by \protect\cite{Georgiev2010} in pink. The diagonal dashed black lines represent $S_{N}$ contours, whereas the horizontal one represents 1 GC.}
	\label{figure:Ngc}
\end{figure*}
\fi%%%%%%%%%%%%%%%%%%%%%%%

\indent We record the following information from the $N_{\mathrm{GC}}$ posterior: The 10th, 50th, and 90th percentiles, the 15.9 \& 84.1 percentiles (i.e. the 1$\sigma$ limits centred on the median). The numbers we quote for $N_{\mathrm{GC}}$ are the median values and the uncertainties span the range of the 1$\sigma$ limits centred on the median; these are the error-bars shown in figure \ref{figure:Ngc}. Note that these estimates are corrected for the colour incompleteness from $\S$\ref{section:coloursel}. Trials with mock datasets showed that the median value is not significantly biased despite the marginal posterior in $f$ being naturally truncated at zero by our model. We find that \nudgzerogc\ out of \nudg\ UDGs have median values of $N_{\mathrm{GC}}$ below one, compared to \nallzerogc\ across the whole sample. However, \nallzerogcsig\ of the whole sample of target galaxies are consistent with having no GCs within $1\sigma$.

\indent Overall, our results show a general increase of $N_{\mathrm{GC}}$ with $\mathrm{M}_{V}$ that is qualitatively consistent with normal dwarf galaxies. While some UDGs are comparable with those of \cite{VanDokkum2017}, most of their objects are quite remarkable when compared to our measurements in terms of having much higher $N_{\mathrm{GC}}$ for a given luminosity. It remains to be seen whether these sources are comparatively rare among LSB galaxies and because Fornax contains less galaxies we see fewer UDGs with GC excess, or that perhaps the increase in environmental density in the Coma cluster plays a positive role in producing such galaxies; this is discussed further in $\S$\ref{section:conclusion}. 

%%%%%%%%%%%%%%%%%%%%%%%%%%%%%%%%%%%%%%%%%%%%%%%%%%%%%%%%%%%%%%%%%% 

\subsection{Colours}
\label{section:colour2}

\begin{figure}
	\includegraphics[width=\linewidth]{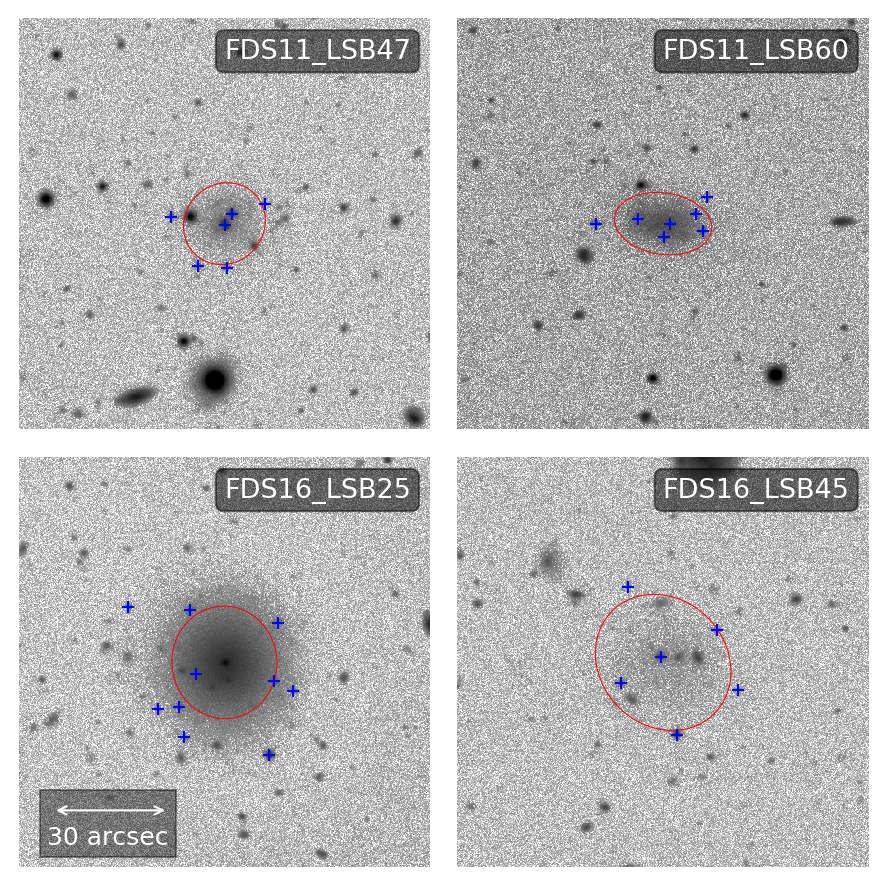}
	\centering
	\caption{$r$-band cut-outs of four of our target galaxies, selected from galaxies with $N_{\mathrm{GC}}\geq$5. The red ellipses represent 1$r_{e,r}$ contours. The blue points identify GCCs with $P_{\mathrm{GC}}\geq0.5$. Of the four galaxies shown, FDS16\_LSB45 is the only one we measure to be large enough to be classified as a UDG. 30$\arcsec$ is $\sim$2.9kpc at Fornax distance.}
	\label{figure:demo2}
\end{figure}

\begin{figure*}
	\includegraphics[width=\linewidth]{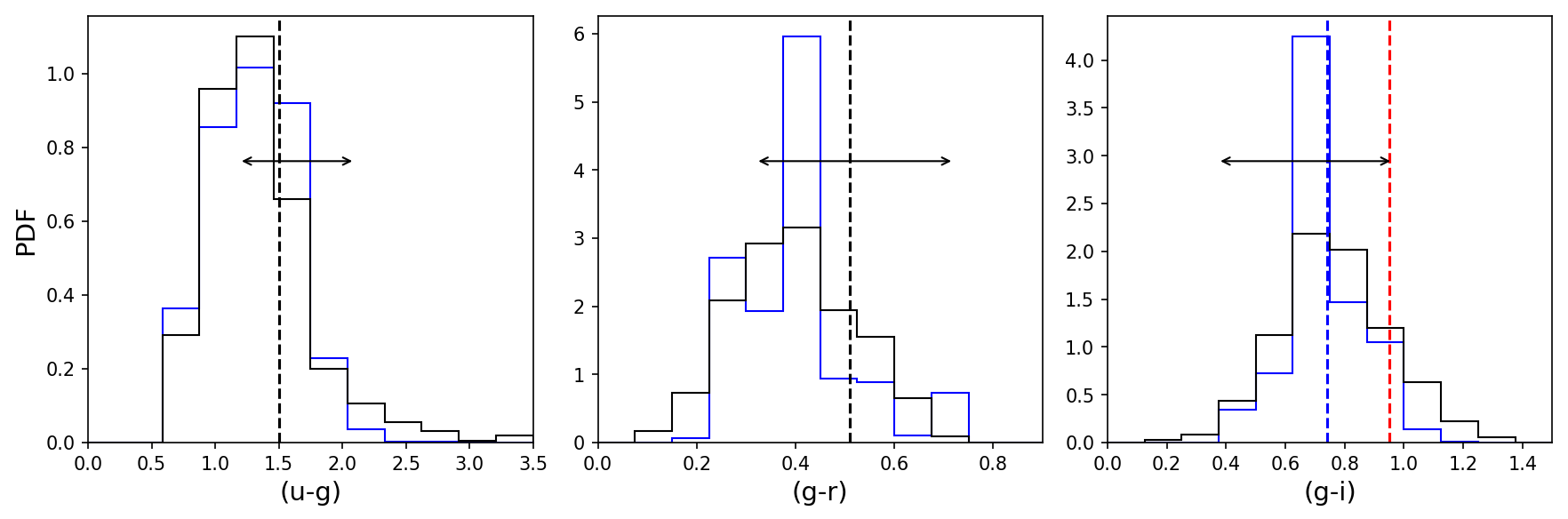}
	\centering
	\caption{Normalised distribution of colours for the GCC sample (black histograms) vs. those of the same sample after being weighted by their probability of belonging to a target galaxy's GC system according to equation \protect\ref{equation:prob} (blue histograms). We note that we select only sources with $m_{g}\leq23$ to overcome the measurement error and ease comparison with \protect\cite{DAbrusco2016}, who used a similar limit. For $(u-g)$ and $(g-r)$, the vertical black dashed lines correspond to the median colours of the sample of spectroscopic sources described in $\S$\ref{section:coloursel}. The dashed blue and red lines in the $(g-i)$ panel correspond to the means of the blue and red components measured by \protect\cite{DAbrusco2016}. The results a consistent with a predominantly blue GC population. The horizontal arrows indicate the $\pm1\sigma$ range of the galaxy colours.}
	\label{figure:colour}
\end{figure*}

Despite already imposing a conservative colour selection criterion in $\S$\ref{section:coloursel}, we can use our results to assess the distribution of colour within the selection box. For each posterior sample, one can assign a probability of belonging to the Plummer profile (i.e. the galaxy) to each GCC  given by

\begin{equation}
P_{\mathrm{GC}}(r)^{j} = \left[1+\frac{2(1-f^{j})\int_{0}^{r_{\mathrm{max}}}\Sigma(r',r_{h}^{j})r'dr')}{f^{j}\Sigma(r,r_{h}^{j})r^{2}}\right]^{-1}
\label{equation:prob}
\end{equation} 

\noindent where $j$ loops over the posterior sample. The result of selecting high-probability GCCs is shown for a selection of galaxies in figure \ref{figure:demo2}. We display the full colour distributions for all our GCCs weighted by their probabilities of cluster membership in figure \ref{figure:colour}. It is clear from these distributions that one-or-two component Gaussian fits are inappropriate, so we limit ourselves to a qualitative discussion based on the weighted histograms.

\indent Comparing the weighted $(g-i)$ histogram with the un-weighted version, it is clear that a narrow peak emerges that is coincident with the blue component measured by \cite{DAbrusco2016} at $<g-i>$=0.74. We conclude that the GC population of our sample is mainly blue. This is consistent with the results of \cite{Peng2006}, who have shown that low luminosity galaxies tend to have predominantly blue GC systems. The blue nature of the GCs is suggestive of young and/or low-metallicity stellar populations. 

\indent In figure \ref{figure:colour} we also show the $\pm1\sigma$ span of the colours of the target galaxies. Clearly the blue peaks we observe in $(g-i)$ and $(g-r)$ are consistent with these colours. In $(g-r)$, the blue peak of the GCs appears shifted to the blue compared to the galaxy colours. However, since this effect is within the $1\sigma$, it is not a significant result.

\subsection{Stellar mass vs Halo mass}

Using our estimates of $N_{\mathrm{GC}}$ together with the empirical trend of \cite{Harris2017} (accounting for the intrinsic scatter in the relation), we are able to estimate the halo mass $M_{halo}$ of the sample of galaxies. For the estimate to be valid, one must assume that $N_{\mathrm{GC}}$ is indeed a reasonable indicator of $M_{halo}$ in the LSB regime. There is limited evidence to support this \citep{Beasley2016a, VanDokkum2017} based on comparisons between $M_{halo}$ measurements inferred from $N_{\mathrm{GC}}$ and those inferred from kinematic measurements. We also estimate the stellar mass using the empirical relation of \cite{Taylor2011} (their equation 8), who used the GAMA survey \citep{Driver2011} to calibrate stellar mass as a function of $g$ and $i$ magnitudes with an intrinsic scatter of $\sim$0.1 dex. 

\begin{figure*}
	\includegraphics[width=\linewidth]{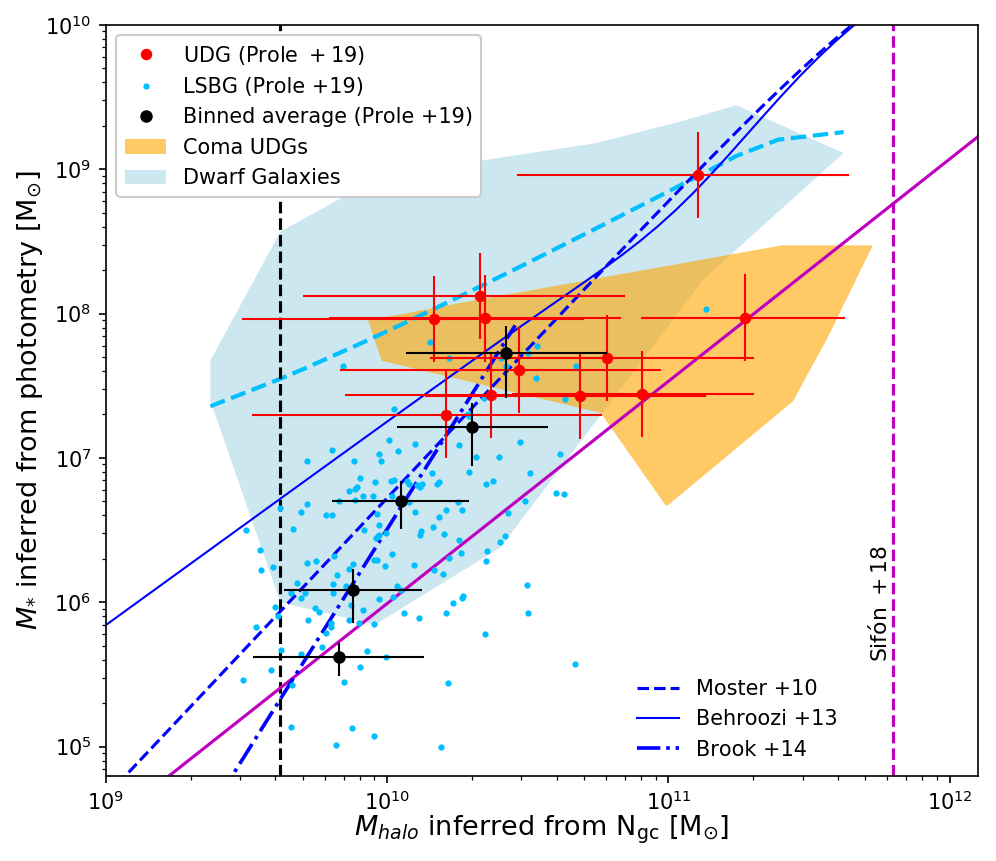}
	\centering
	\caption{Halo mass (derived using $N_{\mathrm{GC}}$) vs. stellar mass. In the figure we show our new measurements, the light blue points correspond to LSBGs and the red to UDGs. The stellar masses for our points were derived from $g$ and $i$ band photometry using the scaling relation of \protect\cite{Taylor2011}. The black error-bars show illustrative binned averages of all our new measurements (calculated in logarithmic bins). The orange region bounds UDG measurements from the literature \protect\citep{VanDokkum2017, Amorisco2018, Lim2018}, We also plot measurements of other ``normal'' dwarf galaxies \protect\citep[][blue region]{Miller2007, Georgiev2010}. The dashed light blue line represents the mean of these sources. The darker blue lines show extrapolated theoretical predictions of \protect\cite{Moster2010}, \protect\cite{Behroozi2013}  \& \protect\cite{Brook14}. The diagonal purple line corresponds to ``enhanced'' GC systems \protect\cite{Amorisco2018}, using the \protect\cite{Harris2017} conversion between $N_{\mathrm{GC}}$ and $M_{halo}$. The dashed purple line indicates the 2$\sigma$ credibility upper-limit on the average mass of UDGs derived from weak lensing \protect\cite{Sifon2018}. Finally, the vertical black dashed line corresponds to $N_{\mathrm{GC}}=1$.}
	\label{figure:sm_hm}
\end{figure*}

\indent We plot our estimates of $M_{halo}$ vs. $M_{*}$ in figure \ref{figure:sm_hm}. We also display other measurements from the literature, including the sample of Coma UDGs from \cite{VanDokkum2017} and the median values measured by \cite{Amorisco2018} (it is worth noting that only three of their sources have $M_{halo}>10^{11}\mathrm{M_{\odot}}$ at 90\% confidence), along with measurements of other dwarf galaxies, including dwarf ellipticals in clusters \citep{Miller2007} as well as late-type dwarfs from a variety of environments including the field \citep{Georgiev2010}. We also show the 2$\sigma$ credibility upper limit on the average mass of UDGs derived from weak lensing by \cite{Sifon2018}, with which our results are consistent. Also we show the extrapolated theoretical predictions from abundance matching of \cite{Moster2010}, \cite{Behroozi2013} and \cite{Brook14}, which were calibrated using observed stellar masses greater than approximately $10^{8}$,  $10^{7}$ and $10^{7} \mathrm{M_{\odot}}$ respectively.

\indent \cite{Forbes2018} show that the $N_{\mathrm{GC}}$ to $M_{halo}$ relation may lose accuracy for $M_{halo}\leq10^{10}\mathrm{M_{\odot}}$, giving systematically higher values of $M_{halo}$ than measured for their sample. According to their study, a better estimator of $M_{halo}$ is the total mass associated with the GC system; however, they note that the assumption of a common mean GC mass is not valid at the low-mass end such that individual GC masses should be measured to get an unbiased estimate of $M_{halo}$, using the empirical relation of \cite{Spitler2009}. While we have not measured the individual GC masses in this work, we note that our estimates of $M_{halo}$ should be considered as upper-limits in light of their result.

\indent Every UDG in our sample is consistent with inhabiting a dwarf sized halo to within 1$\sigma$. There appears to be no significant tendency for UDGs to have enhanced GC populations and therefore enhanced halo mass for their stellar mass. Indeed, there is a qualitatively continuous trend from the LSBGs towards the UDGs.

\indent The overall population is most consistent with an extrapolation of the \cite{Brook14} relation (calibrated with local group dwarf galaxies), but we cannot rule out consistency with that of \cite{Moster2010} or \cite{Behroozi2013} because of the potential for our estimates of $M_{halo}$ to be overestimates. We emphasise however that all models require extrapolation, below stellar masses of 10$^{8}\mathrm{M_{\odot}}$ for the \cite{Moster2010} relation, and 10$^{7}\mathrm{M_{\odot}}$ for that of \cite{Behroozi2013} and \cite{Brook14}.

\indent While no UDGs have estimates of $M_{halo}$ above what might be expected for enriched GC systems  \citep[according to the empirical relation of][]{Amorisco2018}, several of the LSBG sample do show evidence for excess. This might suggest a continuation of GC-enriched systems down to very low stellar mass.

\indent Another point of interest is that our overall sample of LSB galaxies (including UDGs) appears offset from the mean trend of dwarf galaxies, having higher $M_{halo}$ for a given $M_{*}$. While our estimates of $M_{*}$ for the objects from the literature require assumptions about their colours, this may hint that LSB galaxies have systematically higher M/L ratios than normal dwarfs. However this might be a systematic effect; perhaps only LSB galaxies of high M/L ratio are able to survive in the Fornax core.

%\indent However, \cite{Forbes2018} show that the $N_{\mathrm{GC}}$ to $M_{halo}$ relation may lose accuracy for $M_{halo}\leq10^{10}\mathrm{M_{\odot}}$, giving systematically higher values of $M_{halo}$ than measured for their sample. According to their study, a better estimator of $M_{halo}$ is the total mass associated with the GC system, $M_{\mathrm{GC}}$. To compare with our inferences using $N_{\mathrm{GC}}$, we also calculate $M_{halo}$ inferred from $M_{\mathrm{GC}}$ using the \cite{Spitler2009} relation, which \cite{Forbes2018} show can be meaningfully extrapolated to our stellar mass range ($\geq\sim10^{5}\mathrm{M_{\odot}}$). To estimate $M_{\mathrm{GC}}$ for our sample, we use our assumed GCLF along with a GC stellar M/L ratio of 1.88 \cite[i.e. the same as that used by ][for their dwarf galaxies]{Georgiev2010} to calculate an average GC mass of $\sim2\times10^{5}\mathrm{M_{\odot}}$. Scaling this using the inferences on $N_{\mathrm{GC}}$, we arrive at binned results for $M_{halo}$ that are entirely consistent within the uncertainties with what we measure using the \cite{Harris2017} relation.

%%%%%%%%%%%%%%%%%%%%%%%%%%%%%%%%%%%%%%%%%%%%%%%%%%%%%%%%%%%%%%%%%%
\subsection{GC system sizes}

Despite imposing a prior on $r_{h}$ (the GC half number radius) with a mean of 1.5$r_{e}$, we find that our GC systems are typically slightly larger. The median value of $r_{h}$ recovered from the full sample of galaxies is 1.73, with a standard deviation of $\sim0.27$ and range between 0.4 and 2.8 (in units of $r_{e,r}$). We note that the median value of $r_{h}$ for the UDGs is consistent with that of the full sample.

\indent If we use the relation between $r_{h}$ and $M_{halo}$ presented in \cite{Hudson2018}, the resulting $M_{halo}$ estimate is much larger than previously estimated using $N_{\mathrm{GC}}$. For example, for a UDG with $r_{h}$=1.5 kpc should have $M_{halo}$ of around $10^{11.6}\mathrm{M_{\odot}}$, much higher than many of the estimates presented in figure \ref{figure:sm_hm} and generally inconsistent with UDGs with halo mass measurements in the literature \citep[e.g.][]{VanDokkum2017}. While we note that \cite{Hudson2018} make clear that the relation is calibrated only for $M_{halo}\geq10^{12}\mathrm{M_{\odot}}$, we advocate a relation more in-line with that of \cite{Forbes2017} in this regime.

%%%%%%%%%%%%%%%%%%%%%%%%%%%%%%%%%%%%%%%%%%%%%%%%%%%%%%%%%%%%%%%%%%
\subsection{LSBGs vs UDGs}
\label{section:pop}

Now that we have estimates of $N_{\mathrm{GC}}$ for each of our target galaxies, we are in a position to directly compare the LSBG population with the UDGs. The two questions we want to answer are: Does the UDG population show any statistical excess of GCs when compared with the LSBGs in the same luminosity range?; and, is the observed distribution of $N_{\mathrm{GC}}$ for the UDGs discontinuous from that of the LSBGs? 

\begin{figure}
	\includegraphics[width=\linewidth]{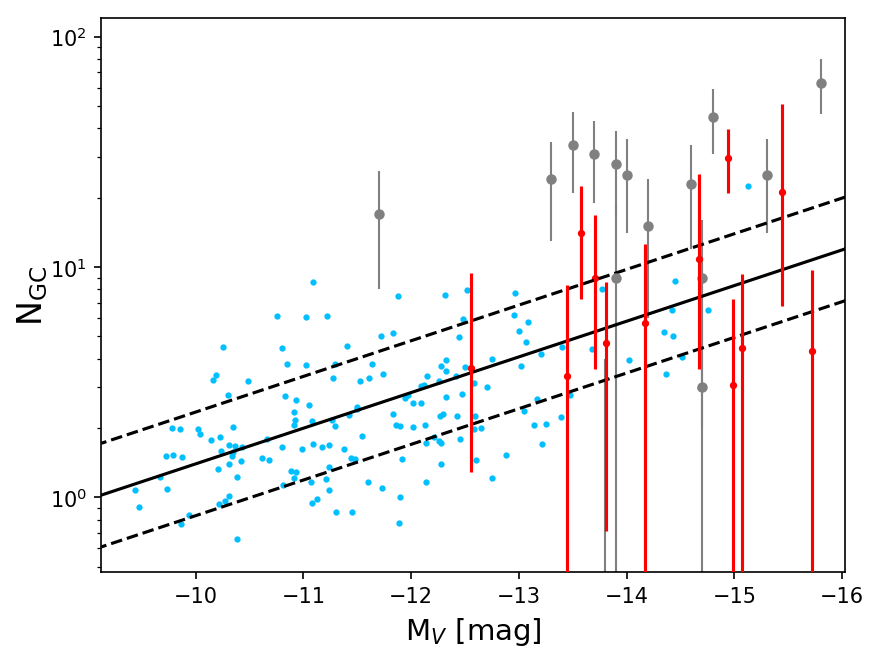}
	\caption{Power-law fit to $N_{\mathrm{GC}}$ vs. M$_{V}$ for our LSBG sample (solid black line) along with an estimate of the 1$\sigma$ scatter (black dashed lines). Blue points: Our LSBG sample. Red errorbars: Our UDG sample. Grey errorbars: The sample of GC-enriched UDGs from \protect\cite{VanDokkum2017}.}
	\label{figure:powerlaw}
\end{figure}

\indent We note that from the appearance of figure \ref{figure:Ngc}, it seems that $N_{\mathrm{GC}}$ vs. $\mathrm{M}_{V}$ can be modelled approximately as a power law. We omit all UDGs from the sample and fit such a relation to our LSBG sample (see also figure \ref{figure:powerlaw}):

\begin{equation}
\bar{N}_{\mathrm{GC}} = (0.04\pm0.02)\times10^{(-0.15\pm0.02)\times \mathrm{M}_{V}}
\label{equation:powerlaw}
\end{equation}

\noindent We note that the scatter in the relation is approximately 0.2 dex across the full magnitude range. Using the fit, we can ask whether our sample of UDGs (ignoring other UDGs from the literature) are consistent with this description. We perform a chi-squared test with the null-hypothesis that the UDGs are drawn from equation \ref{equation:powerlaw}. This results in a $p$-value of 0.30, which means we cannot reject the null hypothesis with an acceptable level of confidence. We therefore conclude that our UDG sample is quantitatively consistent with a continuation of the LSBG sample in this parameter space. We also note that since there is no UDG that has a $N_{\mathrm{GC}}$ measurement convincingly more than 3$\sigma$ above the power-law predicted value, there is no compelling evidence that our UDGs have excessive GC populations.

\indent As as means of comparison, we also do the same test for the population of GC-enriched UDGs from \cite{VanDokkum2017}. While the two tests are not directly comparable since the sample of \cite{VanDokkum2017} is was at-least partially biased to select extreme objects (as in the cases of galaxies DF44 and DFX1), we find that their sample is not consistent with equation \ref{equation:powerlaw}, with a $p$-value much less than 1\%.

\indent  Aside from DF44 and DFX1, the galaxies measured by \cite{VanDokkum2017} also include a list of 12 UDGs selected from the \cite{Yagi2016} catalogue of LSB galaxies that are also present in the Coma Cluster Treasury Program\footnote{https://archive.stsci.edu/prepds/coma/} footprint. Importantly, this should represent a small but unbiased sample of Coma UDGs. After selecting only these sources and repeating the test, we find that the Coma sample is still inconsistent with equation \ref{equation:powerlaw}. This may indicate that UDGs in Coma have more GCs than galaxies in Fornax in the same luminosity range. We find that the choice in prior for the GC half-number radius does not impact this result; for a detailed discussion see appendix \ref{section:prior}.

%%%%%%%%%%%%%%%%%%%%%%%%%%%%%%%%%%%%%%%%%%%%%%%%%%%%%%%%%%%%%%%%%%
\subsection{Effect of the GCLF}
\label{section:GCLF}

\indent As stated in $\S$\ref{section:RE}, we have adopted a Gaussian GCLF with a mean of 24$m_{g}$ and standard deviation of 0.7$m_{g}$. However, dwarf galaxies can have varied GCLFs and it is important to show that our results are robust against this. \cite{Villegas2010} have measured the $g$-band GCLFs for 43 early-type galaxies in the Fornax cluster, down to galaxies with absolute $B$-band magnitudes of around -16. We use this catalogue as a means to test what would happen to our measurements if the GCLF was wider and has turnover magnitude fainter than our adopted value, i.e. to get an upper-limit on the inferences on $N_{\mathrm{GC}}$.

\indent From the \cite{Villegas2010} catalogue, we measure a mean GCLF with mean 24$\pm$0.1$m_{g}$ and a standard deviation of 0.84$\pm$0.21$m_{g}$ after clipping outliers at $2\sigma$. We note that we selected from their catalogue only galaxies with absolute $B$ magnitudes fainter than -18 to target dwarf galaxies for this calculation. This suggests that the GCLF might be wider than what we have assumed previously. Integrating the RE over the $1\sigma$ deeper and wider GCLF and comparing to our previous estimates of the observed GC fraction, we find that the maximum correction in our $N_{\mathrm{GC}}$ is an increase of $\sim$20\%. We find that this is not sufficient to impact or change the overall results of our work (a 20\% increase in $N_{\mathrm{GC}}$ is sufficient to increase a $M_{halo}$ estimate by $\sim$0.1dex).

%%%%%%%%%%%%%%%%%%%%%%%%%%%%%%%%%%%%%%%%%%%%%%%%%%%%%%%%%%%%%%%%%%
\subsection{Effect of Nuclear Star Clusters}
\label{section:NSC}

%\indent  The mass distribution of NSCs is shifted towards the higher masses compared to that of GCs, meaning that are typically $\sim$2 magnitudes brighter. Further, the galaxy nucleation fraction is a function of stellar mass, peaking at 90\% around $M_{*}$=$10^{9}\mathrm{M_{\odot}}$ and decreasing for less (and more) massive galaxies \citep[][]{Sanchez-Janssen2018}. At $M_{*}$$\sim$$10^{5}\mathrm{M_{\odot}}$, essentially no galaxies are expected to host NSCs.

We do not treat potential NSCs any differently from GCs in our analysis; GCCs are defined by their magnitude and colour.  While we have imposed a bright-end magnitude cut on our sample of GCCs, there is still potential for faint NSCs to contaminate our sample and therefore increase the number of GCCs for a target galaxy by one. For the galaxies with low estimates of $N_{\mathrm{GC}}$, this can amount to a significant source of error. However, most of our target galaxies have $M_{*}\leq10^{8}\mathrm{M_{\odot}}$ and are thus expected to have a low nucleation fraction \citep[between 0.7 at $10^{8}\mathrm{M_{\odot}}$ to 0.0 at  $10^{5}\mathrm{M_{\odot}}$, as shown in figure 8 of][]{Sanchez-Janssen2018}.

\indent Removing GCCs close to the centres of galaxies introduces a subjective bias. However, we note that all the galaxies in our sample have already been visually classified as either nucleated or non-nucleated by \cite{Venhola2017}. This number amounts to 10\% of the catalogue. After applying our bright-end magnitude cut, this leaves us with 13 galaxies that are potentially contaminated by a NSC. To quantify the effect this may have on our estimates of $M_{halo}$, we simply drop these sources from the sample and repeat the analysis. We find that the results do not change; the new binned-average estimates of $M_{halo}$ are consistent within much less than $1\sigma$ with those displayed in figure \ref{figure:sm_hm}.

%%%%%%%%%%%%%%%%%%%%%%%%%%%%%%%%%%%%%%%%%%%%%%%%%%%%%%%%%%%%%%%%%%
%%%%%%%%%%%%%%%%%%%%%%%%%%%%%%%%%%%%%%%%%%%%%%%%%%%%%%%%%%%%%%%%%%

\section{Discussion and Conclusions}
\label{section:conclusion}

In this paper we have estimated the halo masses of a sample of \nsources\ LSB galaxies in the \cite{Venhola2017} catalogue using the sizes of their GC populations, including a sub-sample of \nudg\ UDGs. This constitutes the largest sample of low mass galaxies so-far analysed for this kind of study. Candidate globular clusters were identified in the $g$-band using measurements from the \ProFound\ photometry package. We also applied a colour selection based on photometric measurements of a set of spectroscopically confirmed Fornax cluster GCs, using PSF-corrected aperture magnitudes measured in the $u, g, r, i$ bands. Following this, we used a Bayesian Mixture model approach \citep[influenced by the work of][]{Amorisco2018} to infer the total number of GCs associated with each target galaxy, assuming a GCLF appropriate for our sample.

\indent Our estimates of $N_{\mathrm{GC}}$ for the overall population are qualitatively consistent with more compact dwarf galaxies when plotted against $\mathrm{M}_{V}$. We find that the sample of UDGs are statistically consistent with a power-law fit to the $N_{\mathrm{GC}}$ measurements for LSBGs, indicating that there is no discontinuity between the two populations; our sample of UDGs does not have a statistically significant excess of GCs compared to smaller LSB galaxies in the same luminosity range.

\indent We converted the inferences on $N_{\mathrm{GC}}$ to $M_{halo}$ using the empirical relation of \cite{Harris2017}. We additionally derived stellar masses for the galaxies from the empirical relation of \cite{Taylor2011}, using \Imfit\ galaxy models. Overall, the $M_{*}$ estimates are consistent with dwarf galaxies and the $M_{halo}$ estimates are consistent with dwarf sized halos. The LSBG galaxy population appears consistent with the extrapolated \cite{Brook14} abundance-matching relation between $M_{*}$ and $M_{halo}$ and as an extension of measurements from typical dwarf galaxies, but perhaps with slightly larger $M_{halo}$ for the average dwarf at a given $M_{*}$. We suggest that this might be a systematic effect due to the environment; it is possible that only LSB galaxies with high M/L ratios are able to survive in the Fornax core. However, as \cite{Forbes2018} have shown, the $M_{halo}$ estimates may be too large because of a breakdown in accuracy of the  $N_{\mathrm{GC}}$-$M_{halo}$ relation in the low mass regime, and it is not yet clear how this affects our estimates.   

\indent \nudgrich\ of our UDGs have median values of $N_{\mathrm{GC}}$ above the empirical boundary marking GC-rich systems measured by \cite{Amorisco2018}. However, \nudgrichsig\ are consistent within their 1$\sigma$ uncertainties. Several LSBGs also have potential for GC-richness, and  \nlsbrichsig\ are at least 1$\sigma$ above the required threshold. Such objects could make interesting sources for a follow-up study, given that they could represent a continuation of GC-rich objects down to very low stellar mass. If genuine, they could mean that enhanced GC systems are not unique to UDGs and the mechanisms by which UDGs are produced are separate from those by which LSB galaxies gain enriched GC systems, something also observed by \cite{Amorisco2018}.

\indent Using a weighted histogram approach, we have shown that the GC population of our target galaxies is predominantly blue compared to the overall GC population in Fornax. Our result is consistent with the blue peak in $(g-i)$ recorded by \cite{DAbrusco2016}, with a relative depletion of red GCs. Further still, the blue peak of our GC coincides with the $\pm1\sigma$ range of the galaxy colours. There is tentative evidence in $(g-r)$ that the galaxies may be slightly redder than the GCs, but since this is not a significant effect we do not comment on this further. 

%\indent The GCs themselves are well fit by a single component Gaussian model in $(g-i)$, given our observational uncertainties. The mean value of the colour distribution is at $<g-i>=$\mugi, consistent with a predominantly blue population \citep{DAbrusco2016}. This is consistent with the results of \cite{Peng2006}, who have shown that low luminosity galaxies tend to have predominantly blue GC systems. The blue colours suggest that low surface brightness galaxies preferentially have GCs with either low metallicity or young stellar populations. If low surface brightness dwarf galaxies are expected to be part of the primordial population of dwarf galaxies, then it seems more likely that the former is true. We further observe that the $(g-i)$ colours of the galaxies are very similar to that of the GCs, suggesting similar stellar populations.

 \indent The Coma cluster UDGs measured by \cite{VanDokkum2017} seem to have significantly more GCs than what we see in the Fornax cluster. It is notable that our sample is confined to the core of the Fornax cluster. While \cite{Lim2018} show that there is no particular trend of specific frequency $S_{N}$ with cluster-centric radius for bright UDGs in Coma, they also show that $S_{N}$ decreases with cluster-centric radius for fainter galaxies; if anything this could mean that we probe a population with systematically higher $N_{\mathrm{GC}}$ at a given $M_{*}$ than in the cluster outskirts. Two possibilities are that GC-enriched UDGs are comparatively rare objects and we simply do not observe them because Fornax is much less massive than Coma, or the denser environment of the Coma cluster plays a positive role in UDG GC formation or acquisition. We suggest that future studies could provide complete measurements of $N_{\mathrm{GC}}$ for UDGs in other clusters (e.g. Virgo) to address this question. 

\indent Our measurements are sufficient to rule out the failed $L_{*}$ formation theory for UDGs because the halo mass estimates indicate that they reside in dwarf sized halos. We find a continuation in properties between UDGs and smaller LSBGs such that it does not seem that UDGs have a unique or special formation mechanism. Since few of our UDGs are convincingly GC-rich compared to those in Coma, we speculate that this property may be related to environmental density. Perhaps the Coma objects are more efficiently stripped of gas in the Coma core, thus forming fewer stars relative to their halo mass, resulting in systems that appear GC-rich for their stellar mass. A consequence of this effect is that the fraction of GC-rich UDGs should decline with cluster-centric radius, and this may be a valuable way to estimate the relative strengths of secular vs. environmentally-driven formation mechanisms.

%\indent We finally note that we have made our catalogue of GCCs available as an online supplement, which includes the coordinates, corrected photometry as well as the probabilities of being a GC for each of the candidates.

\section*{Acknowledgements} 

We are grateful to Dr. Adriano Agnello for a helpful discussion on Bayesian mixture models. \newline

\noindent We would also like to acknowledge helpful suggestions provided by Dr. Arianna Di Cintio. \newline

\noindent We also thank the referee, Dr. Michael Beasley, for constructive comments. \newline

\noindent C.W. is supported by the Deutsche Forschungsgemeinschaft (DFG, German
Research Foundation) through project 394551440. \newline

\noindent GvdV acknowledges funding from the European Research Council (ERC) under the European Union's Horizon 2020 research and innovation programme under grant agreement No 724857 (Consolidator Grant ArcheoDyn). \newline

\noindent A.V. would like to thank the Vilho, Yrjö, and Kalle Väisälä Foundation of the Finnish Academy of Science and Letters for the financial support. \newline

\noindent R.F.P. and A.V. acknowledge financial support from the European Union's Horizon 2020 research and innovation programme under the Marie Sklodovska-Curie grant agreement No. 721463 to the SUNDIAL ITN network. \newline

\bibliographystyle{mnras}
\bibliography{library.bib}

\appendix

%%%%%%%%%%%%%%%%%%%%%%%%%%%%%%%%%%%%%%%%%%%%%%%%%%%%%%%%%%%%%%%%%%

\section{Galaxy Modelling}
\label{section:imfit}

\begin{figure}
	\includegraphics[width=\linewidth]{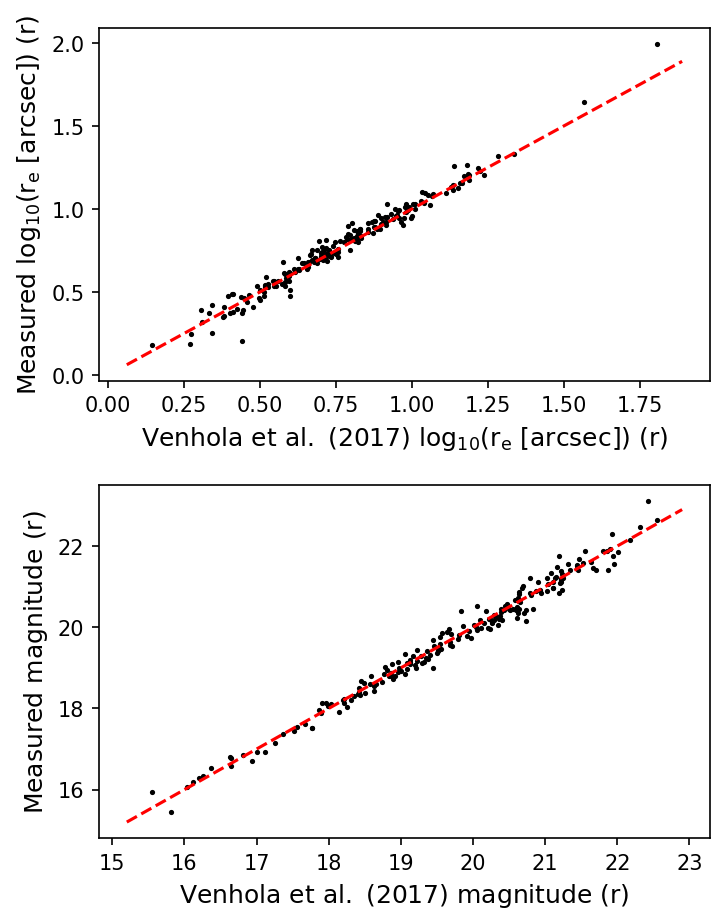}
	\caption{Our galaxy measurements (\Imfit\ ) vs. those of V17 ({\tt GALFIT}). Red dashed line: The one-to-one relation. The RMS of the residuals are 0.18 mag and 0.04 dex in $r_{e}$.}
	\label{figure:imfits}
\end{figure}

\indent We used \Imfit\  to fit single S\'ersic profiles to each target galaxy. Fortunately, \cite{Venhola2017} (hereafter V17)  already provide such fits in the $r$-band. While here we choose to remeasure the profiles for consistency with the other bands, we do make use of these data as initial guesses in the fitting. Our approach was to iteratively fit the galaxy in the $r$-band, each time improving the mask of pixels to ignore in the fit. The general procedure for a galaxy is as follows:

\begin{enumerate}
	\item Obtain an $8\times 8 r^{r}_{e,V17}$ $r$-band cut-out. 
	\item Subtract the V17 model from the result (include nuclear PSF if indicated by V17). 
	\item Use \DeepScan\footnote{https://github.com/danjampro/DeepScan} \citep{Prole2018} to get sky and RMS estimates from the result. 
	\item Create a smoothed image by applying a Gaussian filter with RMS=2 pixels.
	\item Mask all pixels with $\geq 6\sigma$ significance on the smoothed image.
	\item Use \Imfit\  to fit a S\'ersic model to the original data with the sky subtracted, ignoring pixels in the mask.
	\item Repeat steps 2 to 6 three times, each time updating the model image and mask.
	\item Repeat steps 1, 3 \& 6 for the other bands, using the same $r$-band mask in each.
\end{enumerate}

For the \DeepScan\ sky estimates we used a mesh size equal to the image size and performed three masking iterations. If the galaxy was indicated as nucleated by V17, we also fit a Moffat profile simultaneously with the S\'ersic model. We found that in a minority of cases the residuals from the V17 fits were quite large, such that we had to modify the masks manually.

\indent In the case of FDS11\_LSB2, the largest galaxy in our sample (with $r^{r}_{e,V17}=76\arcsec$), we re-binned the data by a factor of 5 to make the fit easier (the original fitting region was $\sim3000\times 3000$ pixels). Over this region the sky background level varies significantly, so we modified the \DeepScan\ sky modelling to use mesh sizes of $\sim 200\arcsec$ and median filtered in $3\times3$ meshes. We note that an image of  FDS11\_LSB2 is displayed in figure 20 of V17.

%\indent We find a small systematic offset between the measurements; ours are generally larger (mean residual of 0.04 dex with residual standard deviation of 0.05 dex) and brighter (mean residual 0.002 mag and residual standard deviation of 0.004 mag). While the origin of this minor discrepancy is not clear, it is likely due to the differences in the background subtraction: 
  
\indent Overall our results are consistent with V17 (figure \ref{figure:imfits}), with a few exceptions. These include FDS11\_LSB2, which we measure to be 1.5 times larger than originally reported. This result was robust against changes in the size of the background mesh. We also note that we measure a slightly lower S\'ersic index $n$ for this object, and $n$ is generally anti-correlated with $r_{e}$. This discrepancy likely arises from the difficulty involved in measuring such a large, diffuse galaxy in a reasonably crowded field with a varying sky; we use {\tt DeepScan} whereas V17 fit a 2D sky plane in their \GALFIT\ \citep{Peng2002} modelling. We also note that V17 did not leave the central coordinate of their model profiles as a free parameter.

\indent Finally, we note that we were not able to obtain stable \Imfit\ models for several sources because they were too faint: FDS12\_LSB42, FDS12\_LSB47, FDS11\_LSB16 \& FDS12\_LSB34. We therefore adopted the fits of V17 for these sources. Since V17 did not measure $(g-i)$ colours, we have omitted them from stellar mass calculations and from figure \ref{figure:sm_hm}.

%%%%%%%%%%%%%%%%%%%%%%%%%%%%%%%%%%%%%%%%%%%%%%%%%%%%%%%%%%%%%%%%%%

\section{Photometric Calibration}
\label{section:photcal}

Starting from the VSTtube-reduced data, we used \ProFound\ to detect and select point sources. We additionally measured fixed-aperture magnitudes for each source with an estimate of that magnitude. These aperture magnitudes had to be corrected for both the limited size of the aperture with respect to the PSF in each band, but also the absolute calibration to AB magnitudes.

\indent While there is no ideal set of standard stars in our footprint with which to calibrate the photometry, Cantiello et al. (in prep) have used a set of existing, overlapping calibrated catalogues (ACSFCS \citep{Jordan2007}; APASS \citep{Henden2012}; SkyMapper \citep{Wolf2018}) to calibrate their photometry in the same data. We have calibrated our own aperture magnitudes by matching our catalogue with theirs, selecting point sources as in $\S$\ref{section:pointsel} and applying a multiplicative correction to our measurements to nullify the mean offset between the measurement pairs. The RMS between our corrected aperture magnitudes with theirs is $\sim$0.05 mag in $g, r, i$ and $\sim0.2$ mag in the shallower $u$-band, for all matching point sources with corrected $g$ magnitudes brighter than 23 mag.

\indent During the calibration it was noticed that the reference catalogue of Cantiello et al. (in prep) contained minor systematic offsets in the stellar locus between individual FDS frames, suggesting a systematic error in the absolute calibration. We have dealt with this by shifting each locus to a common position in colour-colour space. The net result of this is a maximum systematic uncertainty of $\sim\pm0.05$ mag in each colour plane. Finally, we note that since there is currently no available reference catalogue for FDS frame 12 we calibrated the photometry for that frame in accordance with FDS frame 11. This calibration is accurate enough to have negligible effects on our results. 

%%%%%%%%%%%%%%%%%%%%%%%%%%%%%%%%%%%%%%%%%%%%%%%%%%%%%%%%%%%%%%%%%%

\section{Comparison with Miller and Lotz (2007)}
\label{section:ML}

\begin{figure}
	\includegraphics[width=\linewidth]{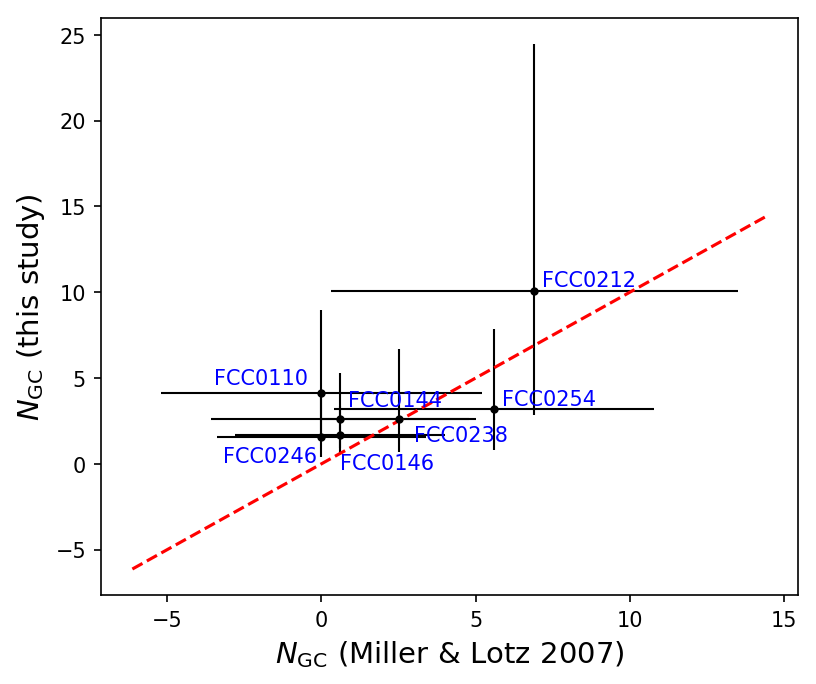}
	\centering
	\caption{Comparison of GC number counts. The black points are median values from the MCMC posterior. The error-bars span the range of $1\sigma$. The red dashed line is the one-to-one relation. A chi-square test (accounting for the errors) shows that the null hypothesis that the measurements are consistent with the one-to-one relation cannot be rejected with confidence greater than around 4\%. }
	\label{figure:ML}
\end{figure}

As several of the sources in the \cite{Venhola2017} catalogue were also identified in the Fornax cluster catalogue \citep{Ferguson1989}, we were encouraged to search for matches in the catalogue of dwarf ellipticals studied by \cite{Miller2007}, who used the HST WFPC2 Dwarf Elliptical Galaxy Snapshot Survey to measure the GC populations for a sample of 69 galaxies. They measured $N_{\mathrm{GC}}$ using apertures of 5 times the exponential scale size of the galaxies, which roughly equates to 3$r_{e}$ for a S\'ersic index $n$=1. 

\noindent We find seven matches: FDS16\_LSB33 (FCC0146), FDS12\_LSB10 (FCC0238), FDS12\_LSB4 (FCC0246), FDS11\_LSB62 (FCC0254), FDS16\_LSB58 (FCC0110), FDS16\_LSB32 (FCC0144) and FDS12\_LSB30 (FCC0212). Overall our results are reasonably consistent (albeit with large error-bars), as is shown in figure \ref{figure:ML}. Note that in the figure one of the sources is not visible because it overlaps with another.

%%%%%%%%%%%%%%%%%%%%%%%%%%%%%%%%%%%%%%%%%%%%%%%%%%%%%%%%%%%%%%%%%%

\section{Choice of prior}
\label{section:prior}

\begin{figure}
	\includegraphics[width=\linewidth]{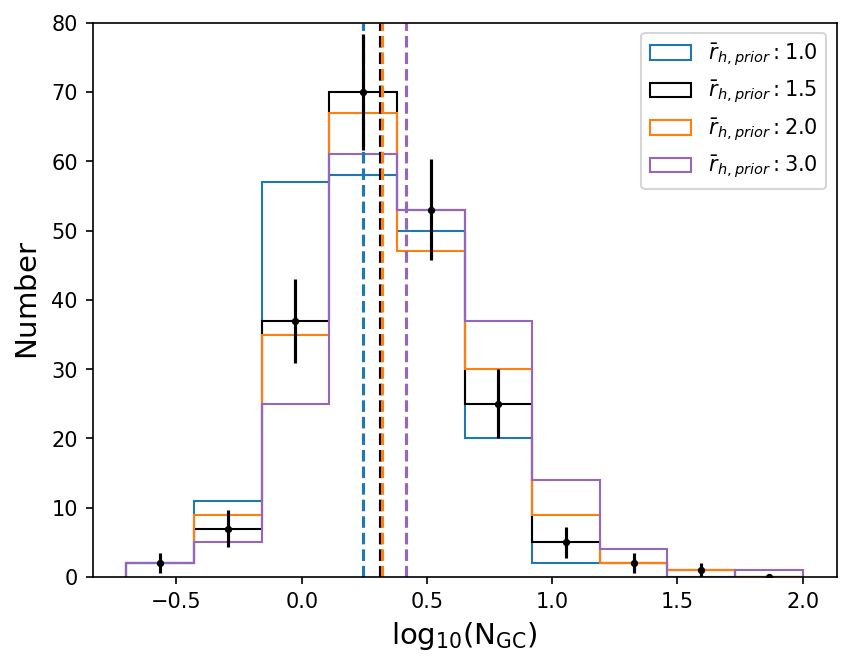}
	\centering
	\caption{The distribution of $N_{\mathrm{GC}}$ estimates for our sample of galaxies as a function on the mean of the prior on the GC half-number radius, $\bar r_{h, prior}$, in units of galaxy half-light radius. Also shown are the median values (dashed lines). Our value of $\bar r_{h, prior}$=1.5$r_{e}$ is adopted from the literature and Poisson error-bars are shown for this value.}
	\label{figure:prior}
\end{figure}

While the choice for the prior on the GC half-number radius $r_{h}$ is justified from previous literature measurements \citep{VanDokkum2017, Amorisco2018, Lim2018}, it is important to show how different choices may affect the results. This is particularly relevant because the spatial distributions of GCs for UDGs is not well known. By running the MCMC using different priors, we show in figure \ref{figure:prior} that, despite the choice of prior in $r_{h}$ strongly influencing the $r_{h}$ posterior, the estimates of $N_{\mathrm{GC}}$ are robust.

\indent A small increase in the mean of the prior on the GC half-number radius, $\bar r_{h, prior}$, is not sufficient to significantly impact our results. However, more dramatic modifications may produce a more pronounced change. In general, lowering $\bar r_{h, prior}$  increases the number of GC-poor systems, while increasing it results in more GC rich systems. However, the median value for the overall population is not significantly altered by using different priors. We finally note that repeating the analysis from $\S$\ref{section:pop} with $\bar r_{h, prior}$=3$r_{e}$ leads us to the same conclusions; the overall result is robust against changes in the prior.

%%%%%%%%%%%%%%%%%%%%%%%%%%%%%%%%%%%%%%%%%%%%%%%%%%%%%%%%%%%%%%%%%%

\onecolumn
\section{Measurements table}
\label{section:table}

\LTcapwidth=\textwidth
\begin{longtable}{ c c c c c c c }
\caption{Results of our analysis. Values enclosed in braces are 10th, 50th and 90th percentiles from the MCMC posterior. Fiducial errors on $M_{V}$ and $M_{*}/\mathrm{M_{\odot}}$ are $\sim0.2$ mag and 0.3 dex respectively. For $r_{e,r}$ the uncertainty is approximately 10\% of the value after measuring the scatter between our measurements and those of \protect\cite{Venhola2017}. We remind the readers that the estimates of $M_{halo}$ should be considered as upper-limits.}\\
Target & M$_{V}$ [mag] & $r_{e,r}$ [kpc] & $\log_{10}\left(M_{*}\ [\mathrm{M_{\odot}}]\right)$ & $r_{h} [r_{e,r}]$ & $N_{\mathrm{GC}}$ & $\log_{10}\left(M_{halo}\ [\mathrm{M_{\odot}}]\right)$  \\ 
\\ \hline \\
FDS10\_LSB2 & -11.0 & 0.46 & 6.3 & $\left\lbrace1.06, 2.04, 2.94\right\rbrace$ & $\left\lbrace0.7, 3.7, 9.4\right\rbrace$ & $\left\lbrace9.5, 10.3, 10.7\right\rbrace$ \\
FDS10\_LSB3 & -9.8 & 0.64 & 6.3 & $\left\lbrace0.57, 1.39, 2.40\right\rbrace$ & $\left\lbrace0.4, 2.0, 4.9\right\rbrace$ & $\left\lbrace9.2, 10.0, 10.4\right\rbrace$ \\
FDS10\_LSB4 & -11.7 & 0.39 & 6.5 & $\left\lbrace0.69, 1.65, 2.66\right\rbrace$ & $\left\lbrace0.2, 1.1, 3.5\right\rbrace$ & $\left\lbrace8.7, 9.7, 10.2\right\rbrace$ \\
FDS10\_LSB5 & -11.1 & 0.94 & 5.6 & $\left\lbrace1.35, 2.28, 3.27\right\rbrace$ & $\left\lbrace2.2, 8.6, 19.6\right\rbrace$ & $\left\lbrace10.0, 10.7, 11.1\right\rbrace$ \\
FDS10\_LSB6 & -10.4 & 0.30 & 5.8 & $\left\lbrace0.77, 1.75, 2.79\right\rbrace$ & $\left\lbrace0.3, 1.4, 3.4\right\rbrace$ & $\left\lbrace9.0, 9.8, 10.2\right\rbrace$ \\
FDS10\_LSB8 & -11.3 & 0.69 & 6.3 & $\left\lbrace0.78, 1.73, 2.73\right\rbrace$ & $\left\lbrace0.3, 2.0, 6.2\right\rbrace$ & $\left\lbrace9.1, 10.0, 10.5\right\rbrace$ \\
FDS10\_LSB9 & -9.9 & 0.35 & 5.8 & $\left\lbrace0.66, 1.37, 2.36\right\rbrace$ & $\left\lbrace0.5, 2.0, 4.6\right\rbrace$ & $\left\lbrace9.3, 10.0, 10.4\right\rbrace$ \\
FDS10\_LSB10 & -11.3 & 0.62 & 6.2 & $\left\lbrace0.64, 1.60, 2.65\right\rbrace$ & $\left\lbrace0.1, 0.9, 2.9\right\rbrace$ & $\left\lbrace8.6, 9.6, 10.1\right\rbrace$ \\
FDS10\_LSB13 & -10.2 & 0.40 & 5.9 & $\left\lbrace0.72, 1.51, 2.52\right\rbrace$ & $\left\lbrace0.4, 1.8, 4.5\right\rbrace$ & $\left\lbrace9.2, 9.9, 10.4\right\rbrace$ \\
FDS10\_LSB14 & -11.1 & 0.50 & 5.6 & $\left\lbrace0.71, 1.69, 2.69\right\rbrace$ & $\left\lbrace0.2, 1.2, 3.6\right\rbrace$ & $\left\lbrace8.9, 9.7, 10.2\right\rbrace$ \\
FDS10\_LSB15 & -11.3 & 0.53 & 6.5 & $\left\lbrace0.43, 1.18, 2.27\right\rbrace$ & $\left\lbrace0.5, 2.2, 5.0\right\rbrace$ & $\left\lbrace9.3, 10.0, 10.4\right\rbrace$ \\
FDS10\_LSB16 & -10.6 & 0.34 & 6.1 & $\left\lbrace0.79, 1.62, 2.54\right\rbrace$ & $\left\lbrace0.3, 1.5, 3.6\right\rbrace$ & $\left\lbrace9.0, 9.8, 10.2\right\rbrace$ \\
FDS10\_LSB23 & -12.6 & 0.78 & 6.7 & $\left\lbrace0.92, 1.91, 2.94\right\rbrace$ & $\left\lbrace0.4, 2.3, 7.2\right\rbrace$ & $\left\lbrace9.1, 10.0, 10.6\right\rbrace$ \\
FDS10\_LSB25 & -14.2 & 2.06 & 7.6 & $\left\lbrace0.85, 1.80, 2.81\right\rbrace$ & $\left\lbrace1.0, 5.7, 15.9\right\rbrace$ & $\left\lbrace9.6, 10.5, 11.0\right\rbrace$ \\
FDS10\_LSB29 & -13.4 & 0.85 & 7.1 & $\left\lbrace0.74, 1.64, 2.61\right\rbrace$ & $\left\lbrace0.4, 2.2, 6.5\right\rbrace$ & $\left\lbrace9.1, 10.0, 10.5\right\rbrace$ \\
FDS10\_LSB35 & -12.3 & 0.35 & 6.6 & $\left\lbrace0.74, 1.70, 2.65\right\rbrace$ & $\left\lbrace0.2, 1.4, 4.0\right\rbrace$ & $\left\lbrace8.9, 9.8, 10.3\right\rbrace$ \\
FDS10\_LSB38 & -12.0 & 0.78 & 6.5 & $\left\lbrace0.19, 1.05, 2.15\right\rbrace$ & $\left\lbrace0.8, 2.8, 6.7\right\rbrace$ & $\left\lbrace9.5, 10.1, 10.5\right\rbrace$ \\
FDS10\_LSB40 & -11.0 & 0.90 & 6.1 & $\left\lbrace1.19, 1.92, 2.73\right\rbrace$ & $\left\lbrace2.1, 6.1, 11.9\right\rbrace$ & $\left\lbrace10.0, 10.5, 10.8\right\rbrace$ \\
FDS10\_LSB41 & -12.3 & 0.63 & 6.8 & $\left\lbrace0.73, 1.64, 2.64\right\rbrace$ & $\left\lbrace0.4, 2.3, 7.8\right\rbrace$ & $\left\lbrace9.1, 10.0, 10.6\right\rbrace$ \\
FDS10\_LSB43 & -11.1 & 0.51 & 6.2 & $\left\lbrace0.71, 1.72, 2.71\right\rbrace$ & $\left\lbrace0.1, 0.9, 3.6\right\rbrace$ & $\left\lbrace8.7, 9.6, 10.2\right\rbrace$ \\
FDS10\_LSB44 & -11.1 & 0.31 & 6.3 & $\left\lbrace0.30, 0.95, 2.15\right\rbrace$ & $\left\lbrace0.5, 1.7, 3.7\right\rbrace$ & $\left\lbrace9.3, 9.9, 10.3\right\rbrace$ \\
FDS10\_LSB45 & -12.1 & 0.53 & 6.7 & $\left\lbrace0.85, 1.84, 2.80\right\rbrace$ & $\left\lbrace0.3, 1.7, 5.6\right\rbrace$ & $\left\lbrace9.0, 9.9, 10.5\right\rbrace$ \\
FDS10\_LSB46 & -10.7 & 0.29 & 5.9 & $\left\lbrace0.66, 1.58, 2.63\right\rbrace$ & $\left\lbrace0.2, 1.4, 4.5\right\rbrace$ & $\left\lbrace8.8, 9.8, 10.3\right\rbrace$ \\
FDS10\_LSB49 & -12.3 & 0.55 & 6.8 & $\left\lbrace0.81, 1.74, 2.79\right\rbrace$ & $\left\lbrace0.3, 1.7, 5.3\right\rbrace$ & $\left\lbrace9.0, 9.9, 10.4\right\rbrace$ \\
FDS10\_LSB51 & -12.1 & 0.80 & 6.8 & $\left\lbrace0.91, 1.81, 2.79\right\rbrace$ & $\left\lbrace0.4, 2.6, 7.2\right\rbrace$ & $\left\lbrace9.2, 10.1, 10.6\right\rbrace$ \\
FDS10\_LSB52 & -13.8 & 1.54 & 7.4 & $\left\lbrace0.41, 1.12, 2.36\right\rbrace$ & $\left\lbrace1.3, 4.7, 10.5\right\rbrace$ & $\left\lbrace9.8, 10.4, 10.8\right\rbrace$ \\
FDS10\_LSB53 & -11.5 & 0.57 & 6.5 & $\left\lbrace0.52, 1.25, 2.31\right\rbrace$ & $\left\lbrace0.4, 1.8, 4.6\right\rbrace$ & $\left\lbrace9.2, 9.9, 10.4\right\rbrace$ \\
FDS10\_LSB54 & -9.7 & 0.33 & 6.7 & $\left\lbrace0.96, 1.82, 2.66\right\rbrace$ & $\left\lbrace0.3, 1.2, 2.6\right\rbrace$ & $\left\lbrace9.0, 9.7, 10.1\right\rbrace$ \\
FDS10\_LSB55 & -12.4 & 0.55 & 7.0 & $\left\lbrace0.88, 1.77, 2.75\right\rbrace$ & $\left\lbrace1.2, 5.0, 11.4\right\rbrace$ & $\left\lbrace9.7, 10.4, 10.8\right\rbrace$ \\
FDS10\_LSB56 & -9.7 & 0.47 & 5.4 & $\left\lbrace0.68, 1.62, 2.65\right\rbrace$ & $\left\lbrace0.2, 1.1, 3.2\right\rbrace$ & $\left\lbrace8.7, 9.7, 10.2\right\rbrace$ \\
FDS11\_LSB4 & -11.2 & 0.46 & 5.9 & $\left\lbrace0.70, 1.77, 2.72\right\rbrace$ & $\left\lbrace0.3, 1.7, 5.5\right\rbrace$ & $\left\lbrace9.0, 9.9, 10.4\right\rbrace$ \\
FDS11\_LSB6 & -10.3 & 0.63 & 6.6 & $\left\lbrace0.53, 1.31, 2.33\right\rbrace$ & $\left\lbrace0.5, 2.0, 5.0\right\rbrace$ & $\left\lbrace9.3, 10.0, 10.4\right\rbrace$ \\
FDS11\_LSB7 & -10.8 & 0.76 & 7.7 & $\left\lbrace1.26, 2.05, 2.92\right\rbrace$ & $\left\lbrace1.6, 6.1, 13.9\right\rbrace$ & $\left\lbrace9.8, 10.5, 10.9\right\rbrace$ \\
FDS11\_LSB8 & -10.3 & 0.48 & 6.2 & $\left\lbrace0.76, 1.71, 2.70\right\rbrace$ & $\left\lbrace0.2, 1.5, 4.5\right\rbrace$ & $\left\lbrace8.9, 9.8, 10.3\right\rbrace$ \\
FDS11\_LSB10 & -11.9 & 0.81 & 6.5 & $\left\lbrace0.53, 1.38, 2.44\right\rbrace$ & $\left\lbrace0.5, 2.1, 5.0\right\rbrace$ & $\left\lbrace9.2, 10.0, 10.4\right\rbrace$ \\
FDS11\_LSB11 & -10.9 & 0.56 & 6.3 & $\left\lbrace0.96, 1.86, 2.76\right\rbrace$ & $\left\lbrace0.4, 2.6, 7.2\right\rbrace$ & $\left\lbrace9.2, 10.1, 10.6\right\rbrace$ \\
FDS11\_LSB13 & -11.1 & 0.48 & 6.3 & $\left\lbrace0.96, 1.97, 2.93\right\rbrace$ & $\left\lbrace0.3, 2.2, 6.3\right\rbrace$ & $\left\lbrace9.0, 10.0, 10.5\right\rbrace$ \\
FDS11\_LSB14 & -11.8 & 1.07 & 7.3 & $\left\lbrace0.84, 1.82, 2.78\right\rbrace$ & $\left\lbrace0.3, 2.3, 7.1\right\rbrace$ & $\left\lbrace9.1, 10.0, 10.6\right\rbrace$ \\
FDS11\_LSB15 & -12.3 & 0.71 & 7.3 & $\left\lbrace0.79, 1.69, 2.65\right\rbrace$ & $\left\lbrace0.7, 3.9, 11.4\right\rbrace$ & $\left\lbrace9.5, 10.3, 10.8\right\rbrace$ \\
FDS11\_LSB16 & -12.5 & 1.50 & -- & $\left\lbrace0.88, 1.79, 2.75\right\rbrace$ & $\left\lbrace0.7, 3.6, 12.0\right\rbrace$ & $\left\lbrace9.4, 10.2, 10.8\right\rbrace$ \\
FDS11\_LSB18 & -10.9 & 0.61 & 5.6 & $\left\lbrace0.08, 0.44, 1.94\right\rbrace$ & $\left\lbrace0.6, 2.2, 5.1\right\rbrace$ & $\left\lbrace9.4, 10.0, 10.4\right\rbrace$ \\
FDS11\_LSB30 & -13.7 & 1.75 & 7.4 & $\left\lbrace1.12, 1.92, 2.79\right\rbrace$ & $\left\lbrace2.2, 8.9, 20.3\right\rbrace$ & $\left\lbrace10.0, 10.7, 11.1\right\rbrace$ \\
FDS11\_LSB35 & -11.3 & 0.69 & 6.0 & $\left\lbrace0.73, 1.45, 2.43\right\rbrace$ & $\left\lbrace0.9, 3.8, 8.4\right\rbrace$ & $\left\lbrace9.6, 10.3, 10.7\right\rbrace$ \\
FDS11\_LSB36 & -10.9 & 0.79 & 6.1 & $\left\lbrace0.74, 1.58, 2.54\right\rbrace$ & $\left\lbrace0.4, 2.3, 6.7\right\rbrace$ & $\left\lbrace9.1, 10.0, 10.5\right\rbrace$ \\
FDS11\_LSB38 & -14.9 & 1.56 & 8.0 & $\left\lbrace0.69, 1.05, 1.53\right\rbrace$ & $\left\lbrace18.0, 29.8, 43.7\right\rbrace$ & $\left\lbrace11.0, 11.3, 11.5\right\rbrace$ \\
FDS11\_LSB39 & -10.2 & 0.38 & 5.0 & $\left\lbrace0.86, 1.74, 2.71\right\rbrace$ & $\left\lbrace0.8, 3.2, 7.3\right\rbrace$ & $\left\lbrace9.5, 10.2, 10.6\right\rbrace$ \\
FDS11\_LSB40 & -9.5 & 0.38 & 4.5 & $\left\lbrace0.65, 1.66, 2.75\right\rbrace$ & $\left\lbrace0.2, 0.9, 2.8\right\rbrace$ & $\left\lbrace8.7, 9.6, 10.1\right\rbrace$ \\
FDS11\_LSB41 & -13.0 & 0.97 & 7.0 & $\left\lbrace0.91, 1.82, 2.82\right\rbrace$ & $\left\lbrace0.4, 2.4, 7.8\right\rbrace$ & $\left\lbrace9.1, 10.0, 10.6\right\rbrace$ \\
FDS11\_LSB42 & -12.1 & 1.22 & 6.5 & $\left\lbrace0.66, 1.65, 2.73\right\rbrace$ & $\left\lbrace0.5, 3.1, 8.8\right\rbrace$ & $\left\lbrace9.3, 10.2, 10.7\right\rbrace$ \\
FDS11\_LSB43 & -9.8 & 0.42 & 5.6 & $\left\lbrace0.84, 1.84, 2.88\right\rbrace$ & $\left\lbrace0.2, 1.5, 4.9\right\rbrace$ & $\left\lbrace8.8, 9.8, 10.4\right\rbrace$ \\
FDS11\_LSB44 & -10.0 & 0.44 & 5.7 & $\left\lbrace0.86, 1.78, 2.74\right\rbrace$ & $\left\lbrace0.3, 1.9, 5.3\right\rbrace$ & $\left\lbrace9.0, 9.9, 10.4\right\rbrace$ \\
FDS11\_LSB45 & -11.4 & 0.71 & 6.3 & $\left\lbrace0.71, 1.78, 2.74\right\rbrace$ & $\left\lbrace0.2, 1.5, 4.9\right\rbrace$ & $\left\lbrace8.8, 9.8, 10.4\right\rbrace$ \\
FDS11\_LSB46 & -10.9 & 0.74 & 6.0 & $\left\lbrace0.78, 1.72, 2.64\right\rbrace$ & $\left\lbrace0.2, 1.3, 4.7\right\rbrace$ & $\left\lbrace8.7, 9.7, 10.4\right\rbrace$ \\
FDS11\_LSB47 & -13.0 & 1.05 & 7.0 & $\left\lbrace0.37, 1.42, 2.38\right\rbrace$ & $\left\lbrace2.5, 7.7, 17.1\right\rbrace$ & $\left\lbrace10.1, 10.6, 11.0\right\rbrace$ \\
FDS11\_LSB49 & -13.7 & 1.34 & 7.4 & $\left\lbrace0.16, 0.71, 2.32\right\rbrace$ & $\left\lbrace1.4, 4.4, 10.2\right\rbrace$ & $\left\lbrace9.8, 10.3, 10.8\right\rbrace$ \\
FDS11\_LSB51 & -10.3 & 0.84 & 6.5 & $\left\lbrace0.82, 1.56, 2.52\right\rbrace$ & $\left\lbrace0.7, 2.8, 6.1\right\rbrace$ & $\left\lbrace9.4, 10.1, 10.5\right\rbrace$ \\
FDS11\_LSB53 & -10.8 & 0.33 & 6.0 & $\left\lbrace1.13, 1.91, 2.78\right\rbrace$ & $\left\lbrace1.1, 3.8, 7.5\right\rbrace$ & $\left\lbrace9.7, 10.3, 10.6\right\rbrace$ \\
FDS11\_LSB55 & -11.4 & 0.89 & 6.0 & $\left\lbrace0.84, 1.78, 2.77\right\rbrace$ & $\left\lbrace0.3, 2.3, 7.1\right\rbrace$ & $\left\lbrace9.1, 10.0, 10.6\right\rbrace$ \\
FDS11\_LSB56 & -11.2 & 0.53 & 6.1 & $\left\lbrace0.62, 1.59, 2.65\right\rbrace$ & $\left\lbrace0.2, 1.1, 3.4\right\rbrace$ & $\left\lbrace8.8, 9.7, 10.2\right\rbrace$ \\
FDS11\_LSB57 & -12.3 & 0.65 & 6.8 & $\left\lbrace0.62, 1.77, 2.71\right\rbrace$ & $\left\lbrace2.3, 7.5, 16.1\right\rbrace$ & $\left\lbrace10.0, 10.6, 11.0\right\rbrace$ \\
FDS11\_LSB58 & -11.2 & 0.90 & 5.9 & $\left\lbrace0.38, 1.72, 2.79\right\rbrace$ & $\left\lbrace1.6, 6.1, 14.4\right\rbrace$ & $\left\lbrace9.8, 10.5, 10.9\right\rbrace$ \\
FDS11\_LSB59 & -13.0 & 0.85 & 7.1 & $\left\lbrace0.34, 1.43, 2.63\right\rbrace$ & $\left\lbrace0.8, 3.7, 10.1\right\rbrace$ & $\left\lbrace9.5, 10.3, 10.7\right\rbrace$ \\
FDS11\_LSB60 & -13.8 & 1.21 & 7.4 & $\left\lbrace0.43, 0.93, 1.89\right\rbrace$ & $\left\lbrace3.6, 8.0, 15.0\right\rbrace$ & $\left\lbrace10.2, 10.6, 10.9\right\rbrace$ \\
FDS11\_LSB61 & -11.9 & 0.61 & 6.6 & $\left\lbrace0.90, 1.73, 2.71\right\rbrace$ & $\left\lbrace0.5, 2.7, 7.2\right\rbrace$ & $\left\lbrace9.3, 10.1, 10.6\right\rbrace$ \\
FDS11\_LSB62 & -14.4 & 1.21 & 7.7 & $\left\lbrace0.91, 1.83, 2.72\right\rbrace$ & $\left\lbrace0.5, 3.4, 10.4\right\rbrace$ & $\left\lbrace9.3, 10.2, 10.8\right\rbrace$ \\
FDS11\_LSB63 & -10.9 & 0.50 & 6.0 & $\left\lbrace0.94, 1.82, 2.71\right\rbrace$ & $\left\lbrace0.4, 2.1, 5.1\right\rbrace$ & $\left\lbrace9.2, 10.0, 10.4\right\rbrace$ \\
FDS11\_LSB64 & -10.8 & 0.30 & 6.2 & $\left\lbrace0.85, 1.73, 2.67\right\rbrace$ & $\left\lbrace0.3, 1.7, 4.0\right\rbrace$ & $\left\lbrace9.1, 9.9, 10.3\right\rbrace$ \\
FDS11\_LSB65 & -12.5 & 0.81 & 6.7 & $\left\lbrace1.22, 2.09, 2.92\right\rbrace$ & $\left\lbrace2.6, 8.0, 16.2\right\rbrace$ & $\left\lbrace10.1, 10.6, 11.0\right\rbrace$ \\
FDS11\_LSB66 & -10.9 & 0.55 & 6.3 & $\left\lbrace0.65, 1.65, 2.68\right\rbrace$ & $\left\lbrace0.2, 1.3, 4.0\right\rbrace$ & $\left\lbrace8.8, 9.7, 10.3\right\rbrace$ \\
FDS11\_LSB67 & -11.7 & 0.55 & 6.4 & $\left\lbrace1.09, 2.00, 2.97\right\rbrace$ & $\left\lbrace1.1, 5.0, 11.8\right\rbrace$ & $\left\lbrace9.7, 10.4, 10.8\right\rbrace$ \\
FDS11\_LSB68 & -12.3 & 0.73 & 6.4 & $\left\lbrace0.79, 1.68, 2.69\right\rbrace$ & $\left\lbrace0.7, 3.7, 9.7\right\rbrace$ & $\left\lbrace9.5, 10.3, 10.7\right\rbrace$ \\
FDS11\_LSB69 & -12.9 & 1.29 & 6.9 & $\left\lbrace0.67, 1.67, 2.65\right\rbrace$ & $\left\lbrace1.6, 6.2, 14.9\right\rbrace$ & $\left\lbrace9.8, 10.5, 10.9\right\rbrace$ \\
FDS11\_LSB71 & -11.6 & 0.55 & 6.0 & $\left\lbrace0.65, 1.72, 2.76\right\rbrace$ & $\left\lbrace0.2, 1.2, 4.0\right\rbrace$ & $\left\lbrace8.8, 9.7, 10.3\right\rbrace$ \\
FDS11\_LSB72 & -11.9 & 0.81 & 6.5 & $\left\lbrace0.60, 1.45, 2.52\right\rbrace$ & $\left\lbrace0.4, 2.0, 5.3\right\rbrace$ & $\left\lbrace9.2, 10.0, 10.4\right\rbrace$ \\
FDS11\_LSB73 & -10.3 & 0.33 & 5.1 & $\left\lbrace0.82, 1.76, 2.76\right\rbrace$ & $\left\lbrace0.3, 1.7, 4.9\right\rbrace$ & $\left\lbrace9.0, 9.9, 10.4\right\rbrace$ \\
FDS11\_LSB74 & -14.0 & 0.97 & 7.3 & $\left\lbrace0.60, 1.42, 2.51\right\rbrace$ & $\left\lbrace0.8, 4.0, 10.0\right\rbrace$ & $\left\lbrace9.5, 10.3, 10.7\right\rbrace$ \\
FDS11\_LSB76 & -10.2 & 0.50 & 5.9 & $\left\lbrace0.70, 1.73, 2.80\right\rbrace$ & $\left\lbrace0.2, 1.3, 4.1\right\rbrace$ & $\left\lbrace8.8, 9.8, 10.3\right\rbrace$ \\
FDS11\_LSB77 & -12.7 & 0.49 & 7.0 & $\left\lbrace0.68, 1.65, 2.61\right\rbrace$ & $\left\lbrace0.2, 1.2, 3.9\right\rbrace$ & $\left\lbrace8.7, 9.7, 10.3\right\rbrace$ \\
FDS11\_LSB78 & -14.3 & 1.19 & 7.6 & $\left\lbrace0.97, 1.93, 2.88\right\rbrace$ & $\left\lbrace1.0, 5.2, 13.8\right\rbrace$ & $\left\lbrace9.6, 10.4, 10.9\right\rbrace$ \\
FDS11\_LSB79 & -12.4 & 0.52 & 6.8 & $\left\lbrace0.29, 1.27, 2.48\right\rbrace$ & $\left\lbrace0.6, 2.3, 5.4\right\rbrace$ & $\left\lbrace9.3, 10.0, 10.4\right\rbrace$ \\
FDS11\_LSB80 & -12.6 & 0.62 & 6.8 & $\left\lbrace0.79, 1.66, 2.65\right\rbrace$ & $\left\lbrace0.3, 2.0, 5.6\right\rbrace$ & $\left\lbrace9.0, 10.0, 10.5\right\rbrace$ \\
FDS11\_LSB81 & -13.1 & 0.78 & 7.1 & $\left\lbrace0.44, 1.35, 2.31\right\rbrace$ & $\left\lbrace2.0, 5.8, 11.7\right\rbrace$ & $\left\lbrace10.0, 10.5, 10.8\right\rbrace$ \\
FDS12\_LSB3 & -13.4 & 1.63 & 7.3 & $\left\lbrace0.88, 1.86, 2.81\right\rbrace$ & $\left\lbrace0.5, 3.4, 11.1\right\rbrace$ & $\left\lbrace9.2, 10.2, 10.8\right\rbrace$ \\
FDS12\_LSB4 & -13.2 & 0.99 & 7.0 & $\left\lbrace0.77, 1.80, 2.77\right\rbrace$ & $\left\lbrace0.2, 1.7, 5.7\right\rbrace$ & $\left\lbrace8.9, 9.9, 10.5\right\rbrace$ \\
FDS12\_LSB5 & -9.9 & 0.47 & 5.0 & $\left\lbrace0.72, 1.84, 2.86\right\rbrace$ & $\left\lbrace0.2, 1.5, 4.9\right\rbrace$ & $\left\lbrace8.9, 9.8, 10.4\right\rbrace$ \\
FDS12\_LSB6 & -9.4 & 0.42 & 5.1 & $\left\lbrace0.74, 1.67, 2.61\right\rbrace$ & $\left\lbrace0.2, 1.1, 3.3\right\rbrace$ & $\left\lbrace8.7, 9.7, 10.2\right\rbrace$ \\
FDS12\_LSB8 & -10.5 & 0.33 & 4.1 & $\left\lbrace1.01, 1.84, 2.68\right\rbrace$ & $\left\lbrace0.9, 3.2, 6.6\right\rbrace$ & $\left\lbrace9.6, 10.2, 10.5\right\rbrace$ \\
FDS12\_LSB9 & -13.4 & 0.99 & 6.8 & $\left\lbrace0.97, 1.91, 2.79\right\rbrace$ & $\left\lbrace0.8, 4.5, 12.8\right\rbrace$ & $\left\lbrace9.5, 10.4, 10.9\right\rbrace$ \\
FDS12\_LSB10 & -13.5 & 1.03 & 6.8 & $\left\lbrace0.89, 1.83, 2.80\right\rbrace$ & $\left\lbrace0.4, 2.8, 9.1\right\rbrace$ & $\left\lbrace9.2, 10.1, 10.7\right\rbrace$ \\
FDS12\_LSB11 & -12.1 & 0.65 & 6.5 & $\left\lbrace0.89, 1.89, 2.86\right\rbrace$ & $\left\lbrace0.3, 2.1, 6.1\right\rbrace$ & $\left\lbrace9.1, 10.0, 10.5\right\rbrace$ \\
FDS12\_LSB12 & -12.1 & 0.93 & 5.9 & $\left\lbrace0.92, 1.84, 2.78\right\rbrace$ & $\left\lbrace0.5, 3.4, 9.7\right\rbrace$ & $\left\lbrace9.3, 10.2, 10.7\right\rbrace$ \\
FDS12\_LSB13 & -13.1 & 1.31 & 6.8 & $\left\lbrace0.98, 1.96, 2.88\right\rbrace$ & $\left\lbrace0.7, 4.7, 14.5\right\rbrace$ & $\left\lbrace9.4, 10.4, 10.9\right\rbrace$ \\
FDS12\_LSB14 & -10.3 & 0.33 & 5.7 & $\left\lbrace0.65, 1.60, 2.63\right\rbrace$ & $\left\lbrace0.2, 1.0, 3.2\right\rbrace$ & $\left\lbrace8.7, 9.6, 10.2\right\rbrace$ \\
FDS12\_LSB16 & -11.5 & 0.73 & 6.1 & $\left\lbrace0.96, 1.85, 2.75\right\rbrace$ & $\left\lbrace0.4, 2.4, 7.3\right\rbrace$ & $\left\lbrace9.2, 10.0, 10.6\right\rbrace$ \\
FDS12\_LSB17 & -11.4 & 0.53 & 6.1 & $\left\lbrace0.79, 1.77, 2.81\right\rbrace$ & $\left\lbrace0.2, 1.6, 5.2\right\rbrace$ & $\left\lbrace8.9, 9.9, 10.4\right\rbrace$ \\
FDS12\_LSB19 & -10.2 & 0.36 & 5.4 & $\left\lbrace0.79, 1.77, 2.80\right\rbrace$ & $\left\lbrace0.3, 1.6, 4.7\right\rbrace$ & $\left\lbrace9.0, 9.8, 10.4\right\rbrace$ \\
FDS12\_LSB20 & -11.8 & 0.49 & 6.5 & $\left\lbrace1.03, 2.00, 2.91\right\rbrace$ & $\left\lbrace1.0, 5.2, 12.2\right\rbrace$ & $\left\lbrace9.6, 10.4, 10.8\right\rbrace$ \\
FDS12\_LSB21 & -12.5 & 0.54 & 6.7 & $\left\lbrace0.96, 1.81, 2.74\right\rbrace$ & $\left\lbrace1.8, 6.0, 12.7\right\rbrace$ & $\left\lbrace9.9, 10.5, 10.9\right\rbrace$ \\
FDS12\_LSB22 & -12.9 & 0.75 & 6.7 & $\left\lbrace0.75, 1.80, 2.83\right\rbrace$ & $\left\lbrace0.2, 1.5, 5.4\right\rbrace$ & $\left\lbrace8.9, 9.8, 10.4\right\rbrace$ \\
FDS12\_LSB23 & -11.2 & 0.50 & 6.0 & $\left\lbrace0.53, 1.38, 2.45\right\rbrace$ & $\left\lbrace0.3, 1.7, 4.5\right\rbrace$ & $\left\lbrace9.1, 9.9, 10.4\right\rbrace$ \\
FDS12\_LSB24 & -10.4 & 0.53 & 6.8 & $\left\lbrace0.76, 1.74, 2.74\right\rbrace$ & $\left\lbrace0.2, 1.6, 5.6\right\rbrace$ & $\left\lbrace8.9, 9.9, 10.5\right\rbrace$ \\
FDS12\_LSB25 & -9.7 & 0.32 & 3.5 & $\left\lbrace0.97, 1.87, 2.84\right\rbrace$ & $\left\lbrace0.3, 1.5, 4.4\right\rbrace$ & $\left\lbrace9.0, 9.8, 10.3\right\rbrace$ \\
FDS12\_LSB26 & -12.3 & 0.48 & 6.3 & $\left\lbrace0.19, 0.91, 2.20\right\rbrace$ & $\left\lbrace0.5, 2.3, 5.4\right\rbrace$ & $\left\lbrace9.3, 10.0, 10.4\right\rbrace$ \\
FDS12\_LSB28 & -12.3 & 0.49 & 6.0 & $\left\lbrace0.57, 1.41, 2.52\right\rbrace$ & $\left\lbrace0.8, 3.5, 9.4\right\rbrace$ & $\left\lbrace9.5, 10.2, 10.7\right\rbrace$ \\
FDS12\_LSB29 & -13.0 & 0.81 & 6.6 & $\left\lbrace0.80, 1.68, 2.67\right\rbrace$ & $\left\lbrace1.2, 5.3, 12.6\right\rbrace$ & $\left\lbrace9.7, 10.4, 10.9\right\rbrace$ \\
FDS12\_LSB30 & -14.7 & 2.00 & 7.7 & $\left\lbrace1.18, 2.22, 3.18\right\rbrace$ & $\left\lbrace2.0, 10.9, 32.2\right\rbrace$ & $\left\lbrace10.0, 10.8, 11.3\right\rbrace$ \\
FDS12\_LSB31 & -10.2 & 0.42 & 5.4 & $\left\lbrace1.11, 1.96, 2.86\right\rbrace$ & $\left\lbrace0.7, 3.4, 8.4\right\rbrace$ & $\left\lbrace9.4, 10.2, 10.7\right\rbrace$ \\
FDS12\_LSB32 & -10.7 & 0.42 & 5.6 & $\left\lbrace0.81, 1.81, 2.79\right\rbrace$ & $\left\lbrace0.3, 1.8, 5.1\right\rbrace$ & $\left\lbrace9.0, 9.9, 10.4\right\rbrace$ \\
FDS12\_LSB33 & -11.2 & 0.61 & 6.1 & $\left\lbrace0.72, 1.68, 2.70\right\rbrace$ & $\left\lbrace0.2, 1.2, 4.1\right\rbrace$ & $\left\lbrace8.8, 9.7, 10.3\right\rbrace$ \\
FDS12\_LSB34 & -11.9 & 1.44 & -- & $\left\lbrace1.25, 2.07, 2.94\right\rbrace$ & $\left\lbrace1.3, 7.5, 20.3\right\rbrace$ & $\left\lbrace9.7, 10.6, 11.1\right\rbrace$ \\
FDS12\_LSB35 & -11.5 & 0.42 & 6.1 & $\left\lbrace0.90, 1.73, 2.65\right\rbrace$ & $\left\lbrace0.2, 1.5, 5.3\right\rbrace$ & $\left\lbrace8.8, 9.8, 10.4\right\rbrace$ \\
FDS12\_LSB42 & -14.5 & 1.25 & -- & $\left\lbrace0.87, 1.83, 2.82\right\rbrace$ & $\left\lbrace0.7, 4.1, 11.9\right\rbrace$ & $\left\lbrace9.4, 10.3, 10.8\right\rbrace$ \\
FDS12\_LSB46 & -11.1 & 0.34 & 5.9 & $\left\lbrace0.70, 1.72, 2.69\right\rbrace$ & $\left\lbrace0.2, 1.0, 3.4\right\rbrace$ & $\left\lbrace8.7, 9.6, 10.2\right\rbrace$ \\
FDS12\_LSB47 & -10.4 & 0.39 & -- & $\left\lbrace0.91, 1.89, 2.87\right\rbrace$ & $\left\lbrace0.3, 1.7, 4.8\right\rbrace$ & $\left\lbrace9.0, 9.9, 10.4\right\rbrace$ \\
FDS12\_LSB50 & -15.0 & 1.55 & 8.0 & $\left\lbrace0.80, 1.72, 2.70\right\rbrace$ & $\left\lbrace0.4, 3.1, 9.6\right\rbrace$ & $\left\lbrace9.2, 10.2, 10.7\right\rbrace$ \\
FDS12\_LSB52 & -12.1 & 0.51 & 6.2 & $\left\lbrace0.94, 1.81, 2.72\right\rbrace$ & $\left\lbrace0.6, 3.1, 8.2\right\rbrace$ & $\left\lbrace9.4, 10.2, 10.6\right\rbrace$ \\
FDS12\_LSB53 & -13.2 & 0.73 & 7.0 & $\left\lbrace0.76, 1.71, 2.73\right\rbrace$ & $\left\lbrace0.3, 2.1, 6.3\right\rbrace$ & $\left\lbrace9.1, 10.0, 10.5\right\rbrace$ \\
FDS12\_LSB54 & -12.4 & 0.46 & 6.6 & $\left\lbrace0.71, 1.74, 2.78\right\rbrace$ & $\left\lbrace0.7, 3.4, 9.1\right\rbrace$ & $\left\lbrace9.5, 10.2, 10.7\right\rbrace$ \\
FDS16\_LSB6 & -12.0 & 0.33 & 6.4 & $\left\lbrace0.56, 1.22, 2.19\right\rbrace$ & $\left\lbrace0.7, 2.0, 4.2\right\rbrace$ & $\left\lbrace9.4, 10.0, 10.3\right\rbrace$ \\
FDS16\_LSB7 & -14.7 & 1.38 & 7.8 & $\left\lbrace0.74, 1.67, 2.80\right\rbrace$ & $\left\lbrace1.5, 6.5, 15.7\right\rbrace$ & $\left\lbrace9.8, 10.5, 11.0\right\rbrace$ \\
FDS16\_LSB10 & -11.9 & 0.40 & 6.5 & $\left\lbrace0.68, 1.65, 2.61\right\rbrace$ & $\left\lbrace0.1, 0.8, 2.4\right\rbrace$ & $\left\lbrace8.6, 9.5, 10.0\right\rbrace$ \\
FDS16\_LSB11 & -14.4 & 1.16 & 7.7 & $\left\lbrace0.81, 1.61, 2.59\right\rbrace$ & $\left\lbrace1.1, 5.0, 12.3\right\rbrace$ & $\left\lbrace9.7, 10.4, 10.8\right\rbrace$ \\
FDS16\_LSB12 & -10.3 & 0.36 & 5.8 & $\left\lbrace0.84, 1.77, 2.75\right\rbrace$ & $\left\lbrace0.2, 1.4, 3.9\right\rbrace$ & $\left\lbrace8.9, 9.8, 10.3\right\rbrace$ \\
FDS16\_LSB14 & -10.1 & 0.46 & 5.9 & $\left\lbrace0.91, 1.81, 2.78\right\rbrace$ & $\left\lbrace0.3, 1.8, 4.7\right\rbrace$ & $\left\lbrace9.0, 9.9, 10.4\right\rbrace$ \\
FDS16\_LSB16 & -9.9 & 0.33 & 5.5 & $\left\lbrace0.65, 1.58, 2.63\right\rbrace$ & $\left\lbrace0.1, 0.8, 2.3\right\rbrace$ & $\left\lbrace8.6, 9.5, 10.0\right\rbrace$ \\
FDS16\_LSB20 & -14.4 & 1.26 & 7.6 & $\left\lbrace0.83, 1.92, 2.88\right\rbrace$ & $\left\lbrace2.4, 8.7, 18.7\right\rbrace$ & $\left\lbrace10.1, 10.7, 11.0\right\rbrace$ \\
FDS16\_LSB24 & -11.0 & 0.40 & 6.1 & $\left\lbrace0.71, 1.64, 2.66\right\rbrace$ & $\left\lbrace0.3, 1.6, 5.1\right\rbrace$ & $\left\lbrace9.0, 9.9, 10.4\right\rbrace$ \\
FDS16\_LSB25 & -15.1 & 1.39 & 8.0 & $\left\lbrace1.12, 1.74, 2.42\right\rbrace$ & $\left\lbrace12.4, 22.5, 36.1\right\rbrace$ & $\left\lbrace10.8, 11.1, 11.4\right\rbrace$ \\
FDS16\_LSB26 & -9.9 & 0.34 & 5.8 & $\left\lbrace0.72, 1.70, 2.68\right\rbrace$ & $\left\lbrace0.1, 0.8, 2.5\right\rbrace$ & $\left\lbrace8.6, 9.5, 10.1\right\rbrace$ \\
FDS16\_LSB28 & -11.9 & 0.49 & 6.6 & $\left\lbrace0.67, 1.67, 2.72\right\rbrace$ & $\left\lbrace0.2, 1.5, 4.7\right\rbrace$ & $\left\lbrace8.9, 9.8, 10.4\right\rbrace$ \\
FDS16\_LSB30 & -10.4 & 0.40 & 5.9 & $\left\lbrace0.80, 1.73, 2.72\right\rbrace$ & $\left\lbrace0.2, 1.2, 3.4\right\rbrace$ & $\left\lbrace8.8, 9.7, 10.2\right\rbrace$ \\
FDS16\_LSB31 & -11.4 & 0.85 & 6.4 & $\left\lbrace1.18, 1.96, 2.83\right\rbrace$ & $\left\lbrace1.2, 4.5, 9.9\right\rbrace$ & $\left\lbrace9.7, 10.4, 10.7\right\rbrace$ \\
FDS16\_LSB32 & -12.5 & 0.63 & 6.8 & $\left\lbrace0.66, 1.53, 2.55\right\rbrace$ & $\left\lbrace0.7, 2.8, 6.7\right\rbrace$ & $\left\lbrace9.4, 10.1, 10.5\right\rbrace$ \\
FDS16\_LSB33 & -12.2 & 0.55 & 6.7 & $\left\lbrace0.25, 0.91, 2.02\right\rbrace$ & $\left\lbrace0.5, 1.8, 4.2\right\rbrace$ & $\left\lbrace9.3, 9.9, 10.3\right\rbrace$ \\
FDS16\_LSB34 & -12.7 & 0.95 & 6.9 & $\left\lbrace0.82, 1.76, 2.71\right\rbrace$ & $\left\lbrace0.6, 3.0, 8.9\right\rbrace$ & $\left\lbrace9.3, 10.2, 10.7\right\rbrace$ \\
FDS16\_LSB35 & -11.6 & 0.47 & 6.5 & $\left\lbrace1.13, 1.97, 2.83\right\rbrace$ & $\left\lbrace0.8, 3.3, 6.6\right\rbrace$ & $\left\lbrace9.5, 10.2, 10.5\right\rbrace$ \\
FDS16\_LSB36 & -13.2 & 0.88 & 7.1 & $\left\lbrace0.76, 1.65, 2.63\right\rbrace$ & $\left\lbrace0.5, 2.7, 7.6\right\rbrace$ & $\left\lbrace9.3, 10.1, 10.6\right\rbrace$ \\
FDS16\_LSB37 & -12.6 & 0.52 & 6.8 & $\left\lbrace0.40, 1.35, 2.42\right\rbrace$ & $\left\lbrace0.8, 3.1, 8.1\right\rbrace$ & $\left\lbrace9.5, 10.2, 10.6\right\rbrace$ \\
FDS16\_LSB38 & -13.2 & 0.88 & 7.0 & $\left\lbrace1.12, 2.12, 3.04\right\rbrace$ & $\left\lbrace0.7, 4.2, 12.7\right\rbrace$ & $\left\lbrace9.4, 10.3, 10.9\right\rbrace$ \\
FDS16\_LSB39 & -10.8 & 0.71 & 5.9 & $\left\lbrace0.54, 1.38, 2.44\right\rbrace$ & $\left\lbrace0.7, 2.8, 6.6\right\rbrace$ & $\left\lbrace9.4, 10.1, 10.5\right\rbrace$ \\
FDS16\_LSB40 & -10.2 & 0.47 & 6.3 & $\left\lbrace1.02, 1.78, 2.69\right\rbrace$ & $\left\lbrace1.3, 4.5, 9.7\right\rbrace$ & $\left\lbrace9.8, 10.4, 10.7\right\rbrace$ \\
FDS16\_LSB41 & -11.4 & 0.49 & 6.4 & $\left\lbrace0.67, 1.66, 2.59\right\rbrace$ & $\left\lbrace0.1, 0.9, 2.9\right\rbrace$ & $\left\lbrace8.7, 9.5, 10.1\right\rbrace$ \\
FDS16\_LSB42 & -12.2 & 0.70 & 6.8 & $\left\lbrace0.79, 1.77, 2.78\right\rbrace$ & $\left\lbrace0.3, 1.8, 5.6\right\rbrace$ & $\left\lbrace9.0, 9.9, 10.5\right\rbrace$ \\
FDS16\_LSB43 & -14.4 & 1.17 & 7.6 & $\left\lbrace0.58, 1.54, 2.61\right\rbrace$ & $\left\lbrace1.9, 6.5, 14.0\right\rbrace$ & $\left\lbrace9.9, 10.5, 10.9\right\rbrace$ \\
FDS16\_LSB44 & -10.0 & 0.67 & 5.1 & $\left\lbrace0.82, 1.67, 2.64\right\rbrace$ & $\left\lbrace0.3, 2.0, 5.9\right\rbrace$ & $\left\lbrace9.1, 10.0, 10.5\right\rbrace$ \\
FDS16\_LSB45 & -13.6 & 1.79 & 7.4 & $\left\lbrace1.04, 1.83, 2.70\right\rbrace$ & $\left\lbrace5.0, 14.0, 26.4\right\rbrace$ & $\left\lbrace10.4, 10.9, 11.2\right\rbrace$ \\
FDS16\_LSB47 & -11.3 & 0.85 & 6.2 & $\left\lbrace1.00, 1.86, 2.79\right\rbrace$ & $\left\lbrace0.6, 3.3, 8.3\right\rbrace$ & $\left\lbrace9.4, 10.2, 10.7\right\rbrace$ \\
FDS16\_LSB49 & -11.5 & 0.81 & 5.9 & $\left\lbrace0.89, 1.80, 2.80\right\rbrace$ & $\left\lbrace0.4, 2.5, 7.6\right\rbrace$ & $\left\lbrace9.2, 10.1, 10.6\right\rbrace$ \\
FDS16\_LSB50 & -13.1 & 0.96 & 7.0 & $\left\lbrace0.93, 1.90, 2.83\right\rbrace$ & $\left\lbrace0.3, 2.1, 6.3\right\rbrace$ & $\left\lbrace9.0, 10.0, 10.5\right\rbrace$ \\
FDS16\_LSB52 & -10.3 & 0.72 & 6.0 & $\left\lbrace0.64, 1.60, 2.68\right\rbrace$ & $\left\lbrace0.2, 1.0, 3.0\right\rbrace$ & $\left\lbrace8.8, 9.6, 10.2\right\rbrace$ \\
FDS16\_LSB54 & -11.2 & 0.66 & 5.7 & $\left\lbrace0.70, 1.55, 2.53\right\rbrace$ & $\left\lbrace0.2, 1.4, 4.0\right\rbrace$ & $\left\lbrace8.9, 9.8, 10.3\right\rbrace$ \\
FDS16\_LSB55 & -12.3 & 0.81 & 6.8 & $\left\lbrace0.73, 1.57, 2.62\right\rbrace$ & $\left\lbrace0.5, 2.7, 7.8\right\rbrace$ & $\left\lbrace9.3, 10.1, 10.6\right\rbrace$ \\
FDS16\_LSB56 & -10.4 & 0.31 & 11.6 & $\left\lbrace0.58, 1.57, 2.60\right\rbrace$ & $\left\lbrace0.1, 0.7, 2.0\right\rbrace$ & $\left\lbrace8.5, 9.4, 10.0\right\rbrace$ \\
FDS16\_LSB58 & -15.1 & 1.70 & 8.0 & $\left\lbrace0.79, 1.56, 2.59\right\rbrace$ & $\left\lbrace1.0, 4.5, 11.4\right\rbrace$ & $\left\lbrace9.6, 10.3, 10.8\right\rbrace$ \\
FDS16\_LSB59 & -10.9 & 0.35 & 6.3 & $\left\lbrace0.75, 1.68, 2.67\right\rbrace$ & $\left\lbrace0.2, 1.2, 3.1\right\rbrace$ & $\left\lbrace8.9, 9.7, 10.2\right\rbrace$ \\
FDS16\_LSB60 & -11.6 & 1.02 & 6.6 & $\left\lbrace0.27, 1.33, 2.44\right\rbrace$ & $\left\lbrace1.2, 3.8, 8.8\right\rbrace$ & $\left\lbrace9.7, 10.3, 10.7\right\rbrace$ \\
FDS16\_LSB63 & -11.9 & 0.64 & 6.7 & $\left\lbrace0.72, 1.73, 2.67\right\rbrace$ & $\left\lbrace0.1, 1.0, 3.4\right\rbrace$ & $\left\lbrace8.7, 9.6, 10.2\right\rbrace$ \\
FDS16\_LSB64 & -12.3 & 1.04 & 6.6 & $\left\lbrace0.96, 1.81, 2.75\right\rbrace$ & $\left\lbrace0.5, 3.2, 9.9\right\rbrace$ & $\left\lbrace9.3, 10.2, 10.7\right\rbrace$ \\
FDS16\_LSB65 & -10.3 & 0.62 & 7.6 & $\left\lbrace0.83, 1.69, 2.68\right\rbrace$ & $\left\lbrace0.3, 1.6, 4.5\right\rbrace$ & $\left\lbrace9.0, 9.8, 10.3\right\rbrace$ \\
FDS16\_LSB66 & -11.0 & 1.04 & 6.8 & $\left\lbrace0.87, 1.75, 2.72\right\rbrace$ & $\left\lbrace0.5, 2.5, 7.0\right\rbrace$ & $\left\lbrace9.2, 10.1, 10.6\right\rbrace$ \\
FDS16\_LSB67 & -10.2 & 0.46 & 5.5 & $\left\lbrace0.68, 1.63, 2.66\right\rbrace$ & $\left\lbrace0.2, 0.9, 2.9\right\rbrace$ & $\left\lbrace8.7, 9.6, 10.1\right\rbrace$ \\
FDS16\_LSB70 & -11.7 & 0.86 & 6.3 & $\left\lbrace0.14, 1.20, 2.42\right\rbrace$ & $\left\lbrace0.9, 3.4, 9.0\right\rbrace$ & $\left\lbrace9.6, 10.2, 10.7\right\rbrace$ \\
FDS16\_LSB71 & -12.7 & 0.64 & 6.9 & $\left\lbrace0.13, 1.34, 2.75\right\rbrace$ & $\left\lbrace1.2, 4.0, 11.4\right\rbrace$ & $\left\lbrace9.7, 10.3, 10.8\right\rbrace$ \\
FDS16\_LSB72 & -12.0 & 0.58 & 6.7 & $\left\lbrace1.04, 2.01, 2.98\right\rbrace$ & $\left\lbrace0.4, 2.6, 7.5\right\rbrace$ & $\left\lbrace9.2, 10.1, 10.6\right\rbrace$ \\
FDS16\_LSB74 & -12.4 & 0.87 & 6.9 & $\left\lbrace0.77, 1.73, 2.79\right\rbrace$ & $\left\lbrace0.3, 1.8, 5.6\right\rbrace$ & $\left\lbrace9.0, 9.9, 10.5\right\rbrace$ \\
FDS16\_LSB75 & -10.8 & 0.53 & 5.8 & $\left\lbrace0.98, 1.87, 2.85\right\rbrace$ & $\left\lbrace1.3, 4.5, 9.7\right\rbrace$ & $\left\lbrace9.7, 10.3, 10.7\right\rbrace$ \\
FDS16\_LSB77 & -12.1 & 0.73 & 6.6 & $\left\lbrace0.65, 1.65, 2.69\right\rbrace$ & $\left\lbrace0.2, 1.2, 4.1\right\rbrace$ & $\left\lbrace8.8, 9.7, 10.3\right\rbrace$ \\
FDS16\_LSB78 & -10.8 & 0.37 & 6.1 & $\left\lbrace0.67, 1.73, 2.77\right\rbrace$ & $\left\lbrace0.2, 1.1, 3.4\right\rbrace$ & $\left\lbrace8.8, 9.7, 10.2\right\rbrace$ \\
FDS16\_LSB79 & -12.6 & 0.88 & 7.1 & $\left\lbrace0.76, 1.72, 2.73\right\rbrace$ & $\left\lbrace0.2, 1.5, 4.9\right\rbrace$ & $\left\lbrace8.9, 9.8, 10.4\right\rbrace$ \\
FDS16\_LSB83 & -12.5 & 0.76 & 6.7 & $\left\lbrace0.90, 1.79, 2.79\right\rbrace$ & $\left\lbrace0.8, 3.7, 9.0\right\rbrace$ & $\left\lbrace9.5, 10.3, 10.7\right\rbrace$ \\
FDS16\_LSB84 & -11.5 & 0.67 & 6.8 & $\left\lbrace0.72, 1.57, 2.60\right\rbrace$ & $\left\lbrace0.5, 3.2, 8.9\right\rbrace$ & $\left\lbrace9.3, 10.2, 10.7\right\rbrace$ \\
FDS16\_LSB85 & -15.7 & 4.23 & 8.1 & $\left\lbrace0.63, 1.66, 2.70\right\rbrace$ & $\left\lbrace0.7, 4.3, 12.5\right\rbrace$ & $\left\lbrace9.5, 10.3, 10.8\right\rbrace$ \\
FDS16\_LSB87 & -12.6 & 0.52 & 6.7 & $\left\lbrace0.88, 1.85, 2.88\right\rbrace$ & $\left\lbrace0.3, 2.0, 6.1\right\rbrace$ & $\left\lbrace9.0, 10.0, 10.5\right\rbrace$ \\
FDS11\_LSB2 & -15.4 & 9.51 & 9.0 & $\left\lbrace1.25, 2.21, 3.22\right\rbrace$ & $\left\lbrace3.3, 21.1, 66.8\right\rbrace$ & $\left\lbrace10.2, 11.1, 11.7\right\rbrace$ \\
FDS10\_LSB27 & -14.7 & 1.45 & 7.8 & $\left\lbrace0.85, 1.78, 2.78\right\rbrace$ & $\left\lbrace0.5, 3.0, 9.4\right\rbrace$ & $\left\lbrace9.3, 10.2, 10.7\right\rbrace$ \\
%\\ \hline \\ 
\label{table:results}
\end{longtable}

\end{document}